\documentclass[aps,preprint,groupedaddress]{revtex4}
\usepackage{amsfonts}
\usepackage{mathrsfs}
\usepackage{graphicx}
\usepackage{subfigure}
\usepackage{setspace}
\usepackage{amsmath}
\usepackage{amssymb}
\usepackage{amsbsy}
\usepackage{bm}
\usepackage{algorithm}
\usepackage{physics}
\usepackage{algpseudocode} 
\newcommand{\half}{\frac12}
\newcommand{\U}{\mathbf{U}}
\newcommand{\I}{\mathbf{I}}
\newcommand{\ra}{\mathbf{r}}
\newcommand{\Sa}{\mathbf{S}}
\begin{document}
\title{Spiking mode-based neural networks}
\author{Zhanghan Lin $^{1}$}
\author{Haiping Huang$^{1,2}$}
\email{huanghp7@mail.sysu.edu.cn}
\affiliation{$^{1}$PMI Lab, School of Physics,
Sun Yat-sen University, Guangzhou 510275, People's Republic of China}
\affiliation{$^{2}$Guangdong Provincial Key Laboratory of Magnetoelectric Physics and Devices, Sun Yat-sen University, Guangzhou 510275, People’s Republic of China}
\date{\today}
\begin{abstract}
Spiking neural networks play an important role in brain-like neuromorphic computations and in studying working mechanisms of neural circuits. One drawback of training a large scale spiking neural network is that updating all weights is quite expensive. Furthermore, after training, all information related to the computational task is hidden into the weight matrix, prohibiting us from a transparent understanding of circuit mechanisms. Therefore, in this work, we address these challenges by proposing a spiking mode-based training protocol, where the recurrent weight matrix is explained as a Hopfield-like multiplication of three matrices: input modes, output modes and a score matrix. The first advantage is that the weight is interpreted by input and output modes and their associated scores characterizing the importance of each decomposition term. The number of modes is thus adjustable, allowing more degrees of freedom for modeling the experimental data. This significantly reduces the training cost because of significantly reduced space complexity for learning. Training spiking networks is thus carried out in the mode-score space. The second advantage is that one can project the high dimensional neural activity (filtered spike train) in the state space onto the mode space which is typically of a low dimension, e.g., a few modes are sufficient to capture the shape of the underlying neural manifolds. We successfully apply our framework in two computational tasks---digit classification and selective sensory integration tasks. Our method thus accelerates the training of spiking neural networks by a Hopfield-like decomposition, and moreover this training leads to low-dimensional attractor structures of high-dimensional neural dynamics. 
\end{abstract}
 \maketitle

\section{Introduction}
Spiking neural activity observed in primates'  brains establishes the computational foundation of high order cognition~\cite{ND-2014}. In contrast to modern artificial neural networks, spiking networks have their own computation efficiency, because the spiking pattern is sparse and also the all-silent pattern dominates the computation. Therefore, studying the mechanism of spike-based computation and further deriving efficient algorithms play an important role in current studies of neuroscience and AI~\cite{Nat-2019}.

In a spiking network (e.g., cortical circuits in the brain), the membrane potential returns to a resting value (hyperpolarization) after a spike is emitted. In an algorithmic implementation, following a spike, the potential is reset to a constant during the refractory period~\cite{ND-2014}. In this short period, the neuron is not responsive to its afferent synaptic currents including external signals. The neural dynamics in the form of spikes is thus not differentiable at spike times, which is in a stark contrast to the rate model of the dynamics, for which a backpropagation through time (BPTT) can be implemented~\cite{Werbos-1990,Elman-1990,Huang-2022}. This nondifferentiable property presents a computational challenge for gradient-based algorithms. A common strategy is to maintain the nondifferentialbe activation in the forward pass, but use a surrogate gradient approach in the backward pass of BPTT~\cite{Zenke-2019}. But previous training of spiking networks is still computational expensive as the full recurrent weight matrix is trained, and thus we propose a mode decomposition learning (explained in detail below) combined with the surrogate gradient to make training spiking networks efficient.  

We review relevant existing methods of training spiking neural networks and highlight the difference from ours. Recurrent neural networks (RNNs) using continuous rate dynamics of neurons can be used to generate coherent output sequences~\cite{Abbott-2009}, in which the network weights are trained by the first-order reduced and controlled error (FORCE) method, where the inverse of the rate correlation matrix is iteratively estimated to update the weights during learning~\cite{Abbott-2009,Huang-2022}. This method can be generalized to supervised learning in spiking neural networks~\cite{Kappen-2016,Force-2017,Kim-2018}. A recent work proposed to train the rate network first and then rescale the synaptic weights to adapt to a spiking setting~\cite{Kim-2019}, while earlier works proposed to  map a trained continuous-variable rate RNN to a spiking RNN model~\cite{Abbott-2016b, Abbott-2016a}. These methods rely either on standard efficient methods like FORCE for rate models, or on heuristic strategies to modify the weights on the rate counterpart. Another recently proposed route is a non-linear voltage-based three-factor learning rule for multilayer networks of deterministic integrate-and-fire neurons~\cite{Zenke-2018}. All these studies involve in real valued weights.
 In contrast, our current mode decomposition framework works directly on the mode space of the recurrent connections underlying the spiking dynamics, for which many biological plausible factors can be incorporated, e.g., membrane and synapse time scales, cell types, refractory time etc. 

Another challenge comes from understanding the computation itself. Even in standard spike-based networks, the weight values on the synaptic connections are modeled by real numbers. When a neighboring neuron fires, the input weight to the target neuron will give a contribution to the integrated currents. With single real weight values, it is hard to dissect which task-related information is encoded in neural activity, especially after learning. In addition, it is commonly observed that the neural dynamics underlying behavior is low-dimensionally embedded~\cite{Ostojic-2021}. There thereby appears a gap between the extremely high dimensional weight space and the actual low dimensional latent space of neural activity. Recent works began to fill the gap using the recurrent rate (rather than spiking) networks, and these works focused on a low-rank decomposition of connectivity matrix (the basis can be used to facilitate a projection)~\cite{Alexis-2022,Ostojic-2023}, or constructing recurrent rate networks to realize specific dynamics on designed manifolds~\cite{Jaz-2020}, or using variants of principal component analysis on recurrent rate dynamics~\cite{Sussillo-2015}.
 To fill the gap in the trained spiking networks, we develop a spiking mode-based neural network (SMNN) here, where the mode vectors constructing the recurrent weights offer the bases for projection of high dimensional neural trajectories. This is not yet addressed in previous studies.

Recent studies focused on the low-rank connectivity hypothesis for recurrent rate dynamics~\cite{Ostojic-2018,Ostojic-2023,Alexis-2022}. The weight matrix in RNNs is decomposed into a sum of a few rank-one connectivity matrices. Recently, this low-rank framework has been applied to analyze the low-rank excitatory-inhibitory spiking networks, with a focus either on mean-field analysis of random networks~\cite{Ostojic-2023b}, decomposition of synaptic input into factor based (partially trained and low-rank) and non-factor based (random) components~\cite{Brian-2023},  or on the capability of approximating arbitrary non-linear input-output mappings using small size networks~\cite{CK-2023}. Therefore, taking biological constraints into account is an active research frontier providing a mechanistic understanding of spike-based neural computations. However, the potential of mode decomposition learning in spiking networks remains unknown, which we shall address in this work.

In our current work, we interpret this kind of low-rank construction \textit{in its full form}, like that in the generalized Hopfield model ~\cite{Jiang-2021,Jiang-2023}. Then a score matrix is naturally introduced to characterize the competition amongst these low-rank modes, where we find a remarkable piecewise power-law behavior for their magnitude ranking. Strikingly, the same mode decomposition learning has been shown to take effects for multi-layered perceptrons trained on a real structured dataset~\cite{Li-2023}, for which the encoding-recoding-decoding hierarchy can be mechanistically explored. As in muliti-layered perceptrons, this full decomposition requires less parameters (a linear space complexity is achievable). 

In essence, we decompose the traditional real-valued weights as three matrices: the left one acts as the input mode space, the right one acts as the output mode space, and the middle one encodes the importance of each mode in the corresponding space. Therefore, we can interpret the weight as two mode spaces and a score matrix. The neural dynamics can thus be projected to the input mode space which is an intrinsically low dimensional space. In addition, the leaky integrated fire (LIF) model can be discretized into a discrete dynamics with intrinsic synaptic dynamics and threshold-type firing, for which a surrogate gradient descent with an additional steepness hyperparameter can be applied~\cite{Zenke-2019,Zenke-2022}.
We then show in experiments how the SMNN framework is successfully applied to classify pixel-by-pixel MNIST datasets and furthermore to context dependent computation in neuroscience. For both computational tasks, we reveal attractor structures in a low-dimensional mode space (in the number of modes) onto which the high dimensional neural dynamics can be projected. Overall, we construct a computational efficient (less model parameters required because of a few dominating modes) and conceptually simple framework to understand challenging spike-based computations (attractor picture in physics). This finding will inspire more physics stuides on the high dimensional spiking neural dynamics including those driven by learning.

\section{Spiking mode-based neural networks}
\label{method}
Our goal is to learn the underlying information represented by spiking time series using SMNNs. This framework is applied to classify the handwritten digits whose pixels are input to the network one by one (a more challenging task compared to the perceptron learning), and is then generalized to the context-dependent computation task, where two noisy sensory inputs with different modalities are input to the network, which is required to output the correct response when different contextual signals are given. This second task is a well-known cognitive control experiment carried out in the prefrontal cortex of Macaque monkeys~\cite{Mante-2013}.

\subsection{Recurrent spiking dynamics with mode-based connections}
Our network consists of an input layer, a hidden layer with $N$ LIF neurons and an output layer. In the input layer, input signals are projected as external currents. Neuron $i$ in the hidden layer (reservoir or neural pool) has an input mode $\bm{\xi}^{\rm in}_i$ and an output mode $\bm{\xi}^{\rm out}_i$, where both modes $\bm{\xi}_i\in\mathbb{R}^{ P}$. All $N$ modes form a pattern matrix $\boldsymbol{\xi}\in\mathbb{R}^{N\times P}$, i.e., each row is a mode vector. The connectivity weight from neuron $j$ to neuron $i$ is then constructed as $W^{\rm rec}_{ij} =\sum_\mu \lambda_\mu\xi^{\rm in}_{i\mu}\xi^{\rm out}_{j\mu}$, where $\lambda_\mu$ is a score encoding the competition among modes, and the connectivity matrix can be written as $\bm{W}^{\rm rec}=\bm{\xi}^{\rm in}\bm{\Sigma}(\bm{\xi}^{\rm out})^\top\in\mathbb{R}^{N\times N}$ where $\bm{\Sigma} \in \mathbb{R}^{P\times P}$ is the importance matrix which is a diagonal matrix $\operatorname{diag}(\lambda_1,\ldots,\lambda_P)$ here. 
For simplicity, we do not take into account Dale's law, i.e., the neuron population is separated into excitatory and inhibitory subpopulations. This biological law could be considered as a matrix product between a non-negative matrix and a diagonal matrix specifying the cell types~\cite{XJ-2016}. 
The activity of the hidden layer is transmitted to the output layer for generating the actual network outputs. Based on SMNNs, the time complexity of learning can be reduced from $\mathcal{O}(N^2)$ to $\mathcal{O}(2NP+P)$. It is typical that $P$ is of the order one, and thus the SMNN has the \textit{linear} training complexity in the network size.

Before specifying the neural dynamics, we remark a few differences from the singular value decomposition method. First, the mode decomposition learning rule is inspired in physics by Hopfield model (a toy model of associative memory~\cite{Hopfield-1982}, see also our two recent papers~\cite{Jiang-2021,Li-2023}). In the generalized form~\cite{Li-2023}, one can flexibly interpret the rank number as the number of patterns, and this number is adjustable. In biology, this decomposition may implement a linear-nonlinear layered computation akin to dendritic integration~\cite{Li-2023}. Second, the elements of the importance matrix is not necessarily positive, and can be learned during the training protocol. Third, the input and output mode vectors are not necessarily orthogonal, allowing for more degrees of freedom to interpret experimental data or execute computational tasks. In addition, these mode vectors help to project the dynamics trajectories in the high-dimensional state space onto the low-dimensional mode space (the dimensionality is determined by $P\ll N$).

In the hidden layer, the membrane potential $\U(t)$ and the synaptic current $\I^{\rm syn}(t)$ of LIF neurons are subject to the following dynamics,
\begin{subequations} \label{condyn}
\begin{align}
\tau_{\rm mem}\frac{\rm{d} \U(t)}{\rm{d}t}
 &=  -\U(t)+\I^{\rm syn}(t)+\I^{\rm ext}(t), \\    
\I^{\rm syn}(t)&= \bm{W}^{\rm rec}\ra(t),
\end{align}
\end{subequations}
where $\tau_{\rm mem}$ is the membrane time constants.  $\I^{\rm ext}$ indicates the external currents. If the recurrent weight matrix does not adopt the proposed decomposition form, and only single weight values are trained, we call this standard type spiking neural network (SNN).  For comparison, we also consider rate networks, where the neural dynamics is specified as follows,
 \begin{equation}\label{rate}
   \tau_{\rm mem}\frac{\rm{d} \U(t)}{\rm{d}t} =  -\U(t)+\bm{W}^{\rm rec}\ra^{\rm rate}(t)+\I^{\rm ext}(t),  
 \end{equation} 
where the firing rate $\ra^{\rm rate}(t)=\rm tanh(\U(t))$. The recurrent connection can take a mode decomposition, and we call this network MDL-RNN, while RNN refers to the standard rate network without mode decomposition. MDL is a shorthand for mode decomposition learning. 

The spikes are filtered by specific types of synapses in the brain. Therefore, we denote $\ra\in \mathbb{R}^{N}$ in Eq.~\eqref{condyn} as the filtered spike train with the following double-exponential synaptic filter~\cite{ND-2014},
\begin{subequations}
\begin{align}
\frac{{\rm{d}}r_i}{{\rm{d}}t}& = -\frac{r_i}{\tau_d}+h_i,\\
\frac{{\rm{d}}h_i}{{\rm{d}}t}& = -\frac{h_i}{\tau_r}+\frac{1}{\tau_r\tau_d}\sum_{t_i^k<t}\delta(t-t_i^k),
\end{align}
\end{subequations}
where $\tau_r$ and $\tau_d$ refer to the synaptic rise time and the synaptic decay time, respectively; $t_i^k$ refers to the $k$-th spiking time of unit $i$.  Different time scales of synaptic filters are due to different types of receptors (such as fast AMPA, relatively fast GABA, and slow NMDA receptors) with different temporal characteristics~\cite{ND-2014}. Note that the above synaptic filter has also been used in a previous work~\cite{Kim-2019}, which proposed to train the rate network first and then rescale the synaptic weights to adapt to the spiking setting. We remark that other types of synaptic filters can also be used, e.g. the simple exponential type used in previous works~\cite{Force-2017,Zenke-2022}.

By definition,
\begin{equation}
\I^{\rm ext}(t) = \bm{W}^{\rm in}\mathbf{u}(t)
\end{equation}
where the time-varying inputs $\mathbf{u}\in\mathbb{R}^{N_{\rm in}}$ are fed to the network via $\bm{W}^{\rm in}\in\mathbb{R}^{N\times N_{\rm in}}$. Corresponding to the continuous dynamics in Eq.~\eqref{condyn} ,
the membrane potential is updated in discrete time steps by~\cite{Cramer-2019}
\begin{equation}
\U[n+1] = \Big( \lambda_{\rm mem}\U[n]+(1-\lambda_{\rm mem})\I[n])\Big)\odot \Big( \mathcal{I}-\Sa[n] \Big),
\end{equation}
with $n$ is the time step, $\odot$ denotes the element-wise product, $\mathcal{I}$ is an all-one vector of dimension $N$, and the membrane decay factor $\lambda_{\rm mem}\equiv\exp(-\frac{\Delta t}{\tau_{\rm mem}})$. $\Delta t$ is a small time interval or step size for solving the ordinary differential equations [Eq.~\eqref{condyn}].  We show details of derivations in Appendix~\ref{app-1}. A spike thus takes place at a time measured in the unit of $\Delta t$.
$\Sa(t)$ is the associated spiking output of neuron $i$, computed as $S_i(t)=\Theta(U_i(t)-U_{\rm thr})$ with a spike threshold $U_{\rm thr}$ (set to one in the following analysis, i.e., $U_{\rm thr}=1$) and Heaviside step function $\Theta$. We also consider the refractory period whose length is denoted by $t_{\rm ref}$. The factor $\mathcal{I}-\Sa[n]$ resets the membrane potential to zero after a spike event (i.e., reset to the resting potential $U_{\rm res}=0$). If the refractory period is considered, the membrane potential will stay at zero for a short duration, e.g., $t_{\rm ref}=2\,{\rm ms}$. In the following, $\I[n]$ denotes the total afferent synaptic currents at the time step $n$ and is calculated as
\begin{equation}
\I[n+1] = \bm{W}^{\rm rec}\ra[n]+\bm{W}^{\rm in}\mathbf{u}[n].
\end{equation}

\begin{figure}
\centering
\subfigure[]{
\parbox[][4cm][c]{0.75\linewidth}{
  \includegraphics[width = \linewidth]{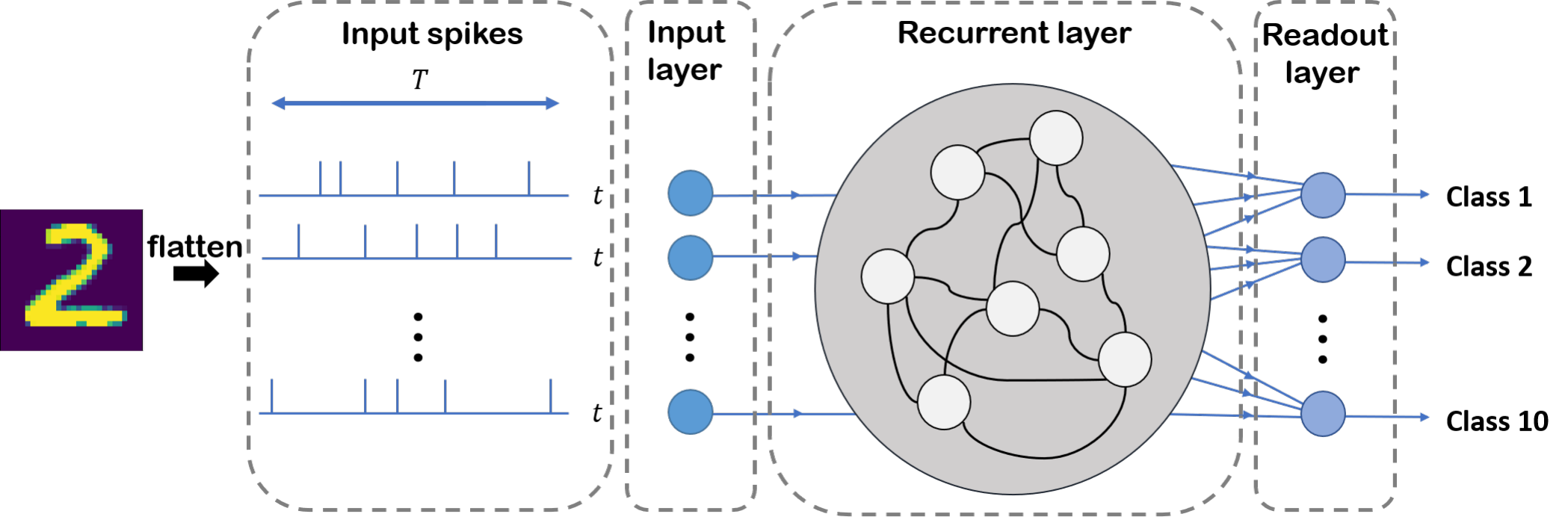}}
}
\subfigure[]{
\parbox[][4cm][c]{0.2\linewidth}{
  \includegraphics[width = \linewidth]
  {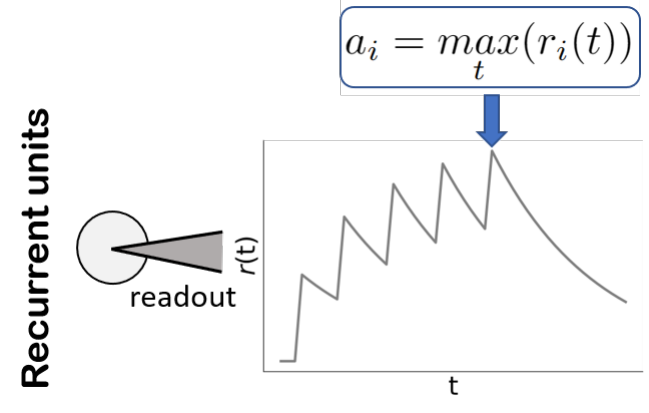}}
}
\subfigure[]{
  \includegraphics[width = 0.95\linewidth]{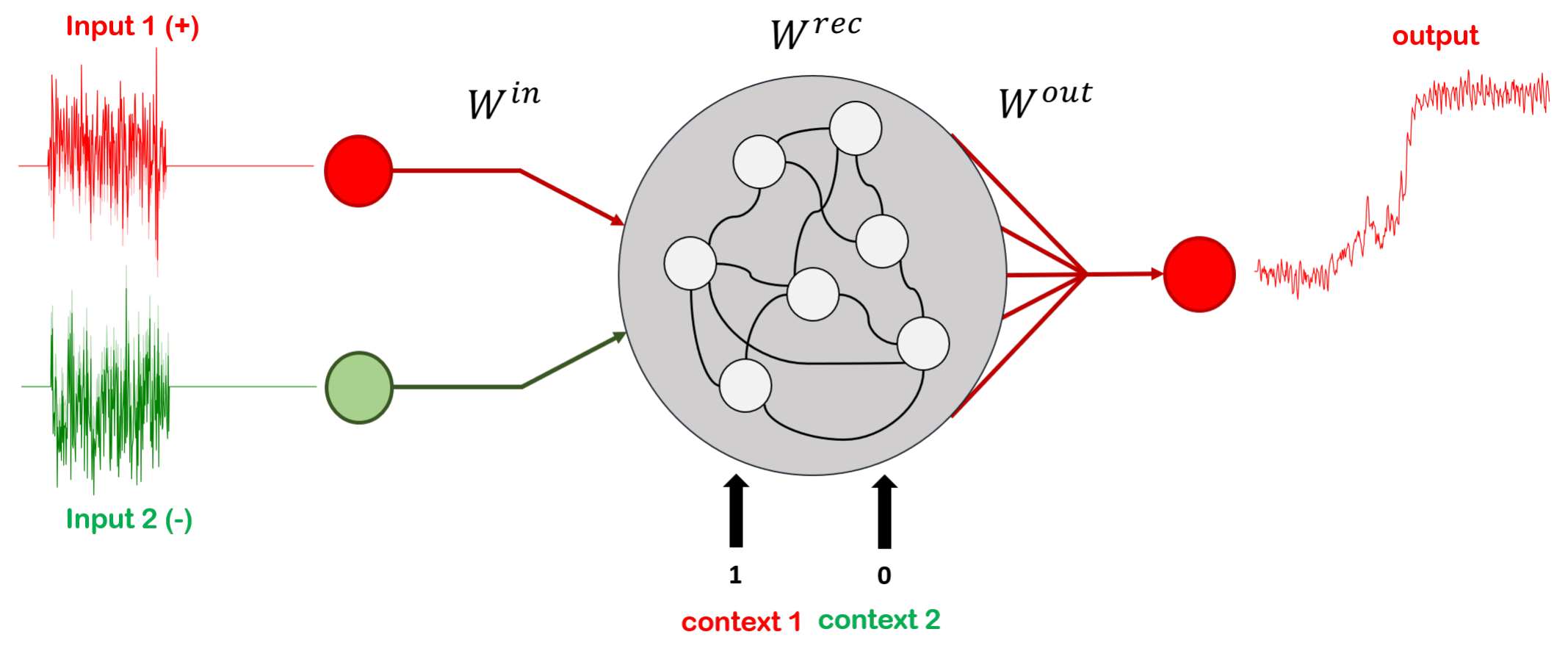}
}
\caption{Model structures for MNIST and contextual integration tasks. (a) Model structure for MNIST task. Each image is converted to a spiking activity input to the recurrent reservoir, by Poisson spiking neurons whose rate is determined by the pixel intensity (see details in the main text). $T=20\, {\rm{ms}}$ in our learning setting. (b) The activity profile of one readout unit for the MNIST task. Only the maximal value is taken for classification. (c) Model structure for contextual integration task. If the cued (context 1 or context 2) input signal is generated using a positive offset value, then the network is supervised to produce an output approaching $+1$ regardless of the irrelevant input signals (e.g., those coming from the other context).}\label{model}
\end{figure}

\subsection{Computational tasks}
\subsubsection{MNIST data learning}
\label{mnistexp}
The MNIST dataset consists of handwritten digits, each of which is composed of 28x28 pixels, belonging to $10$ different classes (6000 images for each class).
The dataset is commonly used as a benchmark classification for neural networks~\cite{mnist}. The handwritten digit is a static image, and thus must be transformed into spiking time series.  We thus introduce an additional transformation layer consisting of spiking neurons before the hidden layer. More precisely, each image is converted into spiking activity using one neuron per pixel. The transformation layer is then modeled by a population of Poisson spiking neurons with a maximum firing rate $f_{\rm max}$. Each neuron in this layer generates a Poisson spike train with a rate $f_i=\frac{g_i}{255}f_{\rm max}$, where $g_i$ is the corresponding pixel intensity.  We present the same static image every $0.2\, {\rm{ms}}$ (step size) and for a total of $100$ time steps as an input to the transformation layer (therefore for a duration $20\ {\rm{ms}}$), and this input image is converted into spiking activity as described above to the hidden layer [Figure~\ref{model} (a) and (b)].

\subsubsection{Context-dependent learning task}
It is fundamentally important for our brain to attend selectively to one feature of noisy sensory input, while the other features are ignored. The same modality can be relevant or irrelevant depending on the contextual cue, thereby allowing for flexible computation (a fundamental ability of cognitive control). The neural basis of this selective integration was found in the prefrontal cortex of Macaque monkeys~\cite{Mante-2013}. In this experiment, monkeys were trained to make a decision about either
the dominant color or motion direction of randomly moving colored dots. Therefore, the color or motion indicates the context for the neural computation. This context-dependent flexible computation can be analyzed by training a recurrent rate neural network~\cite{Mante-2013,XJ-2016,Miconi-2017}.
However, a mode-based training of spiking networks is lacking so far. 

Towards a more biologically plausible setting, we train a recurrent spiking network using our SMNN framework. The network has two sensory inputs of different modalities in analogy to motion and color, implemented as a Gaussian trajectory whose mean is randomly chosen for each trial, but the variance is kept to one. In addition, two contextual inputs (cues indicating which modality should be attended to) are also provided. The network is then trained to report whether the sensory input in the relevant modality has a positive mean or a negative mean (offset). Within this setting, we hypothesize that four attractors would be formed after learning, corresponding to left motion, right motion, red color and green color (see Table~\ref{tab-main}), as expected from the Monkey experiments~\cite{Mante-2013}. We would test this hypothesis using our SMNN framework.
In simulations, we consider $500$ time steps with a step size of $0.2\,{\rm{ms}}$, and noisy signals are only present during the stimulus window (from $10\, {\rm{ms}}$ to $50\, {\rm{ms}}$) [see an illustration in Figure~\ref{model} (c) or Figure~\ref{ctx} (c)].
\begin{table}[bt]
\centering
\caption{\label{tab-main} Attractor type corresponding to the monkey's experiment}
\begin{tabular}{lll}
\hline
  Offset &Contextual input & Attractor type\\
  \hline
   positive & (1,0) & left attractor\\
     negative & (1,0) & right attractor\\
       positive & (0,1) & red attractor\\
        negative & (0,1) & green attractor\\
\hline
\end{tabular}
\end{table}

\section{Simulation results of mode decomposition learning}
\label{exp}
Throughout our experiments, the Heaviside step function is approximated by a sigmoid surrogate \textit{only} in the backward pass of the learning process
\begin{equation}\label{sur}
\Theta(x) \approx \frac{x}{1+k|x|},
\end{equation}
where a steepness parameter $k$ = 25. This is because, $\lim_{k\to\infty}\Theta'(x)=\delta(x)$, where $x=U-U_{\rm thr}$. In the forward pass, the original form of the step function is used to model the spiking process. Other parameters include $\Delta t=0.2\,{\rm{ms}}$, $\tau_{\rm mem}=20\,{\rm{ms}}$, $\tau_{d}=30\,{\rm{ms}}$, $\tau_r=2\,{\rm{ms}}$, and the refractory period $t_{\rm ref}=2\,{\rm{ms}}$ for all tasks. These hyper-parameters are summarized in Table~\ref{tab2} (see Appendix~\ref{app-3}). Because we have not imposed sparsity constraint on the connectivity, we take the initialization scheme that $[\bm{\xi}^{\rm in}\bm{\Sigma}(\bm{\xi}^{\rm out})^\top]_{ij}\sim O(\frac{1}{\sqrt{N}})$, similar to what is done in multi-layered perceptron learning~\cite{Li-2023}. 
The training is implemented by the adaptive-moment-estimation (Adam~\cite{adam}) based stochastic gradient descent algorithm that minimizes the loss function (cross-entropy for classification or mean-squared error for regression in the second task) with a learning rate of $0.001$. We remark that training in the mode space can be done by using automated differentiation on PyTorch. However, we still leave the detailed derivation of the learning rule to Appendix~\ref{app-2}. Codes are available in our GitHub~\cite{ZH-2023}.
\subsection{MNIST classification task}
As a proof of concept, we first apply the SMNN to the benchmark MNIST dataset.
The maximum of recurrent unit activity over time is estimated by $a_i = \operatorname{max}\limits_{t}\{r_i(t)\}$~\cite{Zenke-2022}, which are read out by the readout neurons as follows,
\begin{equation}
\mathbf{o} = {\operatorname{softmax}}(\bm{W}^{\rm out}\mathbf{a}),
\end{equation}
where the readout weight matrix $\bm{W}^{\rm out}\in\mathbb{R}^{10\times N}$. Using BPTT, the gradients in the mode space $\bm{\theta}=(\bm{\xi}^{\rm in},\bm{\Sigma},\bm{\xi}^{\rm out})$  is calculated as follows,
\begin{equation}
\begin{aligned}
\frac{\partial\mathcal{L}}{\partial \bm{\xi}^{\rm in}}&=\sum_{t=1}^T\frac{\partial\mathcal{L}}{\partial \I(t)}\frac{\partial\I(t)}{\partial \bm{W}^{\rm rec}}\frac{\partial\bm{W}^{\rm rec}}{\partial \bm{\xi}^{\rm in}}=\sum_{t=1}^T\frac{\partial\mathcal{L}}{\partial \I(t)}\ra(t-1)\bm{\xi}^{\rm out}\bm{\Sigma},\\
\frac{\partial\mathcal{L}}{\partial \lambda_\mu}&=\sum_{t=1}^T\frac{\partial\mathcal{L}}{\partial \I(t)}\frac{\partial\I(t)}{\partial \bm{W}^{\rm rec}}\frac{\partial\bm{W}^{\rm rec}}{\partial \lambda_\mu}=\sum_{t=1}^T\frac{\partial\mathcal{L}}{\partial \I(t)}\ra(t-1)\bm{\xi}^{\rm in}_\mu(\bm{\xi}^{\rm out}_\mu)^\top,\\
\frac{\partial\mathcal{L}}{\partial \bm{\xi}^{\rm out}}&=\sum_{t=1}^T\frac{\partial\mathcal{L}}{\partial \I(t)}\frac{\partial\I(t)}{\partial \bm{W}^{\rm rec}}\frac{\partial\bm{W}^{\rm rec}}{\partial \bm{\xi}^{\rm out}}=\sum_{t=1}^T\frac{\partial\mathcal{L}}{\partial \I(t)}\ra(t-1)\bm{\xi}^{\rm in}\bm{\Sigma},
\end{aligned}
\end{equation}
where the $\mu$-th column of the pattern matrix $\boldsymbol{\xi}$ is denoted by $\bm{\xi}_\mu^{\rm in/out}\in\mathbb{R}^N$, $T$ is the length of the training trajectory, $\mathcal{L}$ is the loss function, whose gradients $\frac{\partial\mathcal{L}}{\partial\I(t)}$ with respect to the total synaptic current are calculated explicitly in Appendix~\ref{app-2}.  Each training mini-batch contains $100$ images, and LIF networks with different mode sizes ($P = 1, 2, \ldots 50, 100$) are trained. 

\begin{figure}
\centering

\subfigure[]{
  \includegraphics[width = 0.31\linewidth]{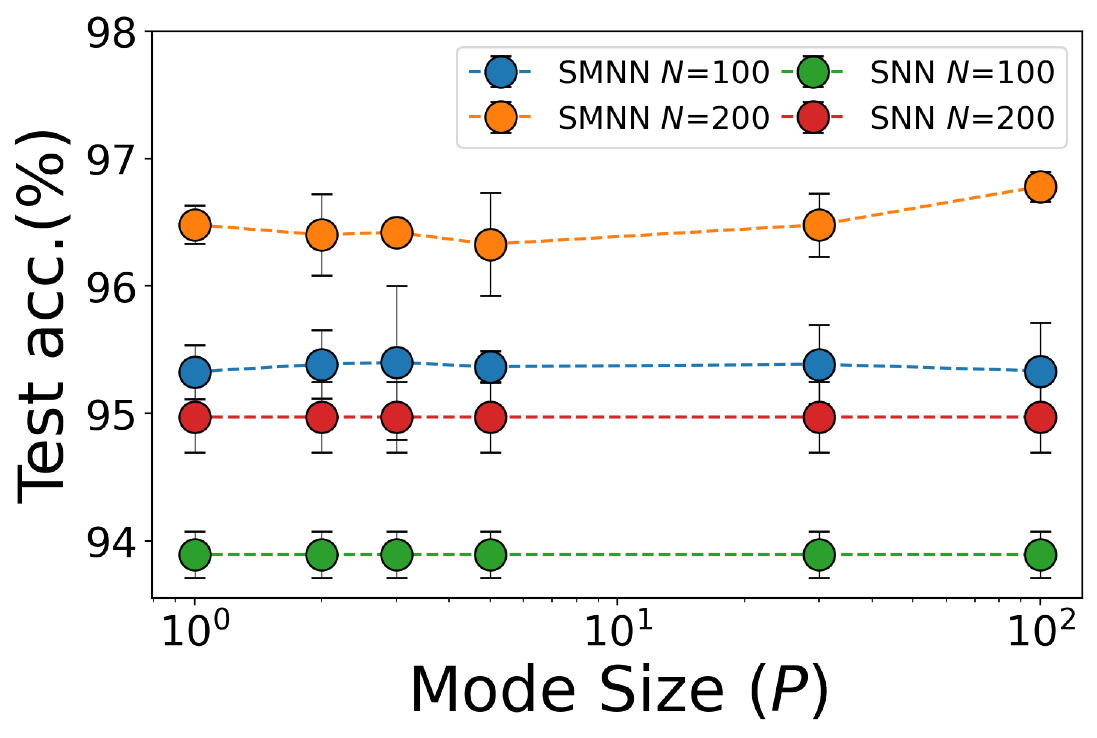}
}
\subfigure[]{
  \includegraphics[width = 0.31\linewidth]{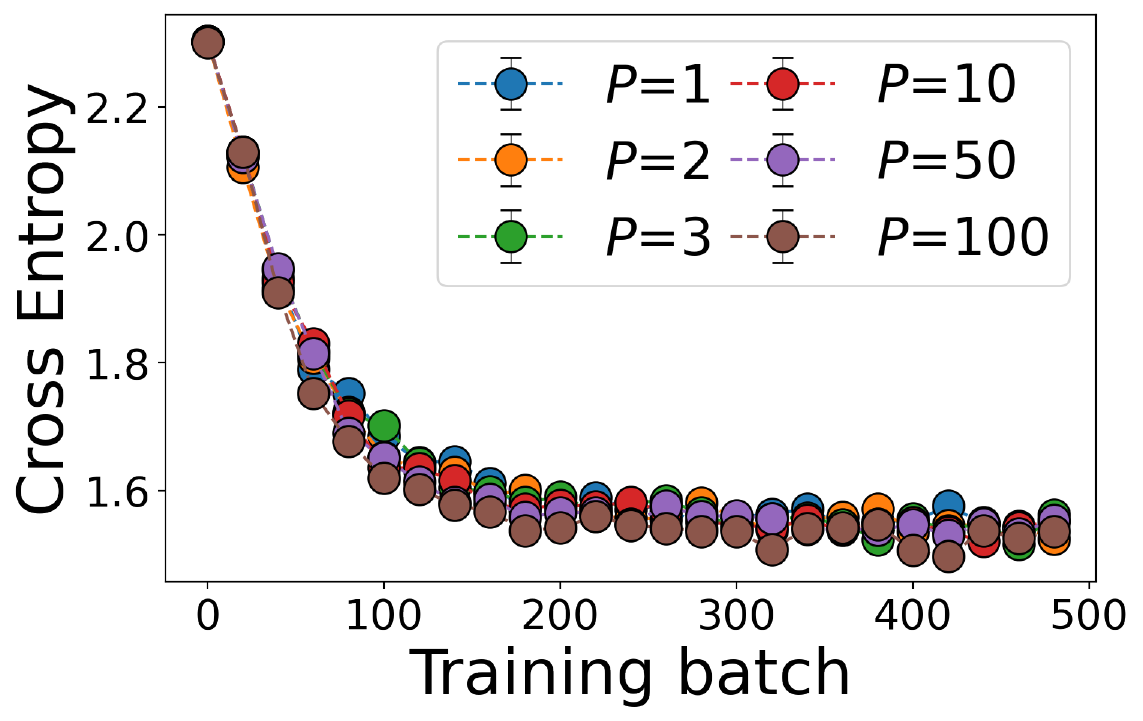}
}
\subfigure[]{
  \includegraphics[width = 0.31\linewidth]{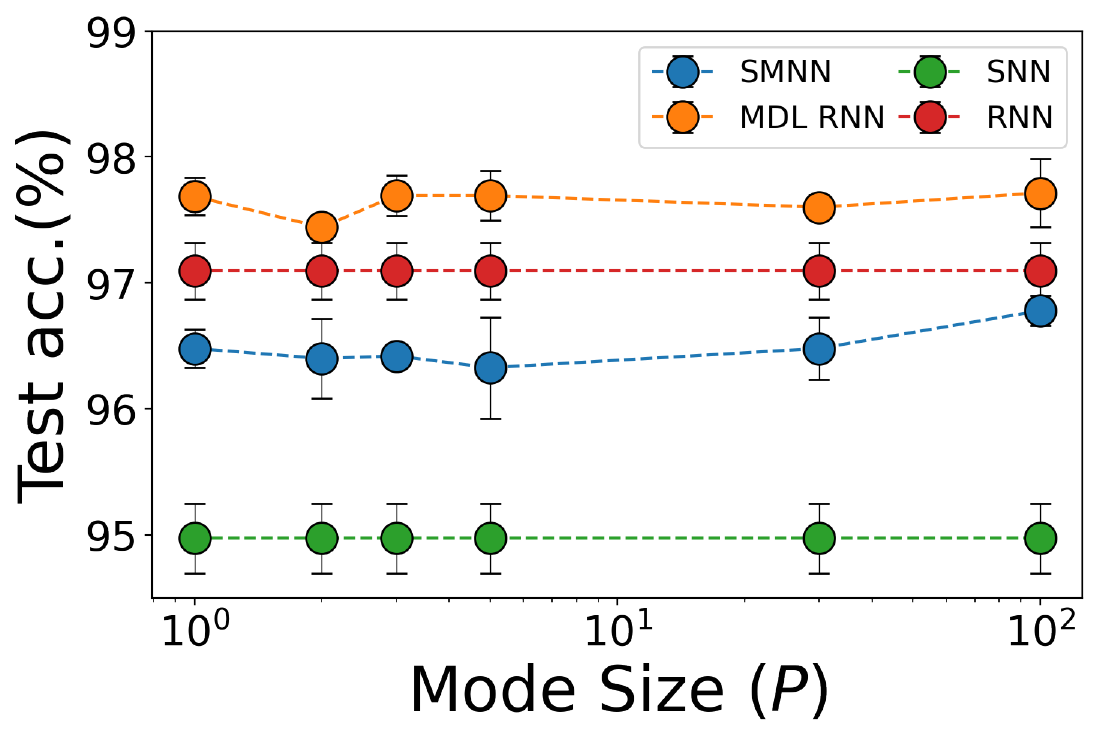}
}
\subfigure[]{
  \includegraphics[width = 0.45\linewidth]{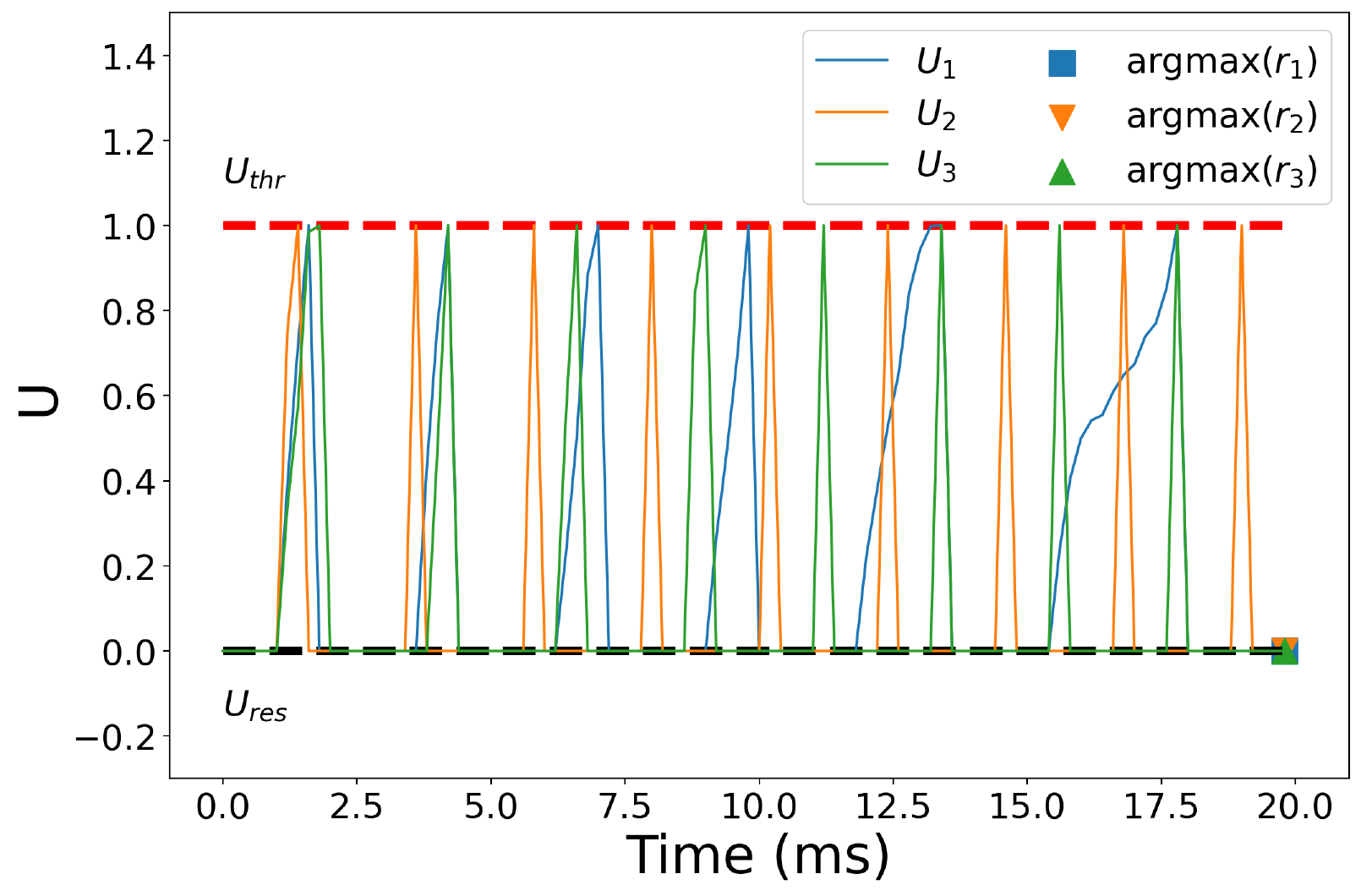}
}
\subfigure[]{
  \includegraphics[width =0.45 \linewidth]{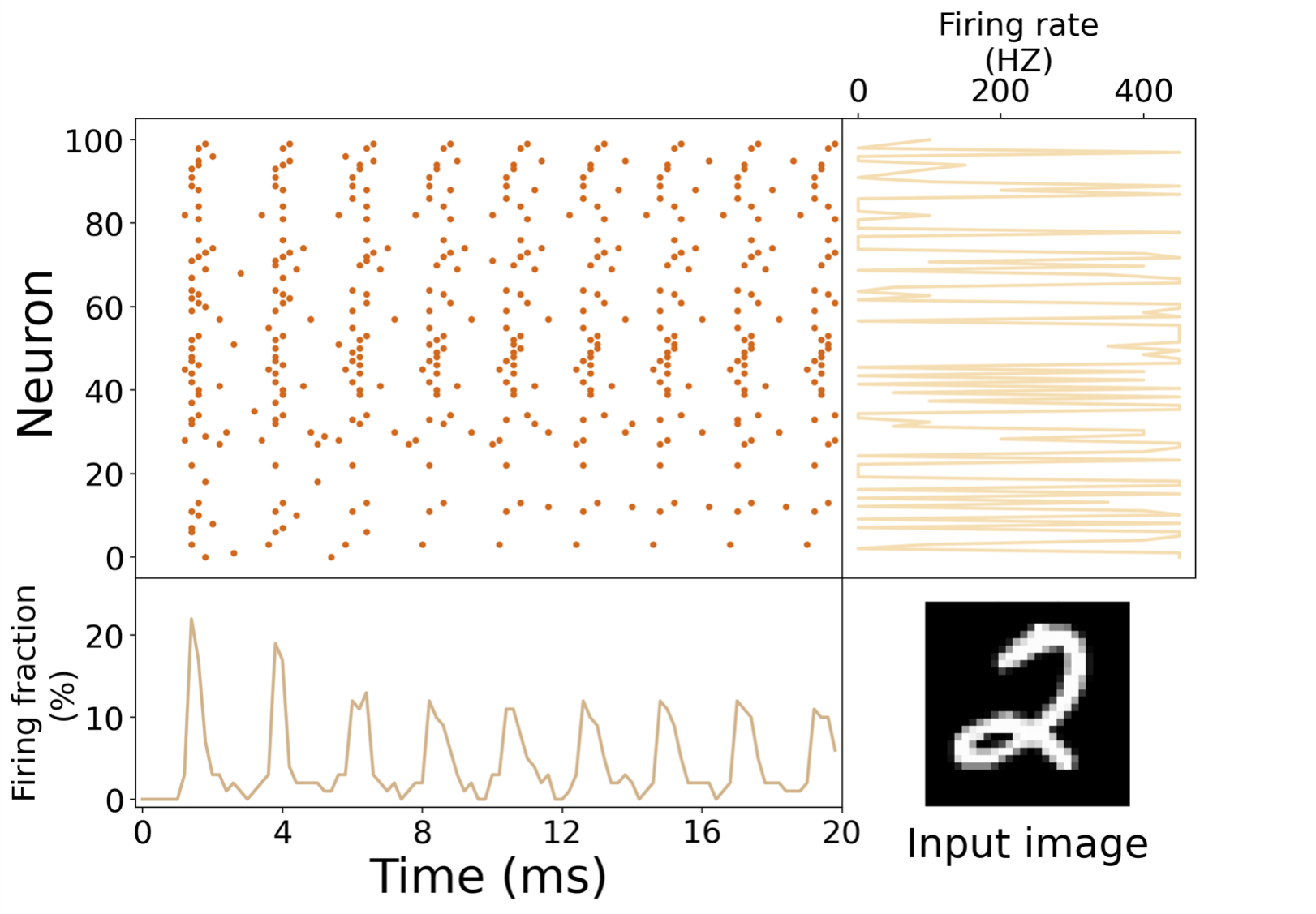}
}
  \caption{Learning performance of MNIST classification task. Five independent runs are used to estimate the standard deviation. (a) Test accuracy versus different mode size. Spiking model without mode decomposition learning (this counterpart is called SNN) is compared. SMNN indicates the mode-based learning of spiking networks. (b) Loss function as a function of training mini-batch. Each mini-batch is composed of $100$ digit images. The mode size varies, and the network size $N=100$.
  (c) Comparison of the accuracies between rate and spiking models with the same fixed network size $N$=200. Rate networks with mode-decomposition learning (MDL RNN) and without MDL (RNN) are also considered. (d) Membrane potential traces for three typical reservoir neurons in response to spike train inputs after training.  The triangle or square marks when the filtered spike train takes a maximum value. $(P,N)=(3,100)$. Note that the displayed three moments for the maximal firing rates overlap with each other,  but this is not the case for other neurons. (e) Spike trains of reservoir neurons with neuron firing rate (right) and population firing fraction (bottom) in response to an input image of digit $2$. $(P,N)=(3,100)$.}\label{digit}
\end{figure}

With increasing mode size, the test accuracy increases with a small margin [Figure~\ref{digit} (a) and (b)]. A better generalization is achieved by a larger neural population. The network size is a key parameter to determine the performance. Surprisingly, even for the smallest mode size ($P = 1$), the network could be trained to perform well, reaching an accuracy of about $96.5\%$ for $N=200$. In comparison with other network counterparts, the SMNN performs much better than SNN, demonstrating that the mode decomposition plays an important role as well [Figure~\ref{digit} (c)]. In addition, the rate network is slightly better than SMNN for this digit classification task, while the SMNN bears a low population firing rate as well as fewer parameters and is thus energetically efficient and fast. However, for the rate network counterpart, the mode-decomposition learning is still a key element to boost the performance [Figure~\ref{digit} (c)].
We also plot the membrane potential profile for a few representative neurons in a well-trained network, and observe a network oscillation [Figure~\ref{digit} (d) and (e)], which reflects a neural population coding of a repeated presentation of the same digit in our stimulus setting (see Sec~\ref{mnistexp}). 

\subsection{Contextual integration task}
In the contextual integration task, the network activity is read out by an affine transformation as
\begin{equation}
o(t) = \bm{W}^{\rm out}\ra(t),
\end{equation}
where $\bm{W}^{\rm out}\in\mathbb{R}^{1\times N}$ refers to the readout weights, and $\ra(t)$ is the filtered spike train. We use the root mean squared (RMSE) loss function defined as 
\begin{equation}\label{manteloss}
\mathcal{L} = \sqrt{\sum_{t = 0}^T[z(t)-o(t)]^2},
\end{equation}
where $z(t)$ is the target output in time $t$, taking zero except in the response period, and $T$ denotes the time range.
In analogy to the previous MNIST classification task, the gradient in the mode-score space $\bm{\theta}=(\bm{\xi}^{\rm in},\bm{\Sigma},\bm{\xi}^{\rm out})$ for the current context-dependent computation could be similarly derived (see details in Appendix~\ref{app-2}). The learning rate was set to $0.001$, and each training mini-batch contains $100$ trials. 
\begin{figure}
\centering
\subfigure[]{
  \includegraphics[width = 0.3\linewidth]{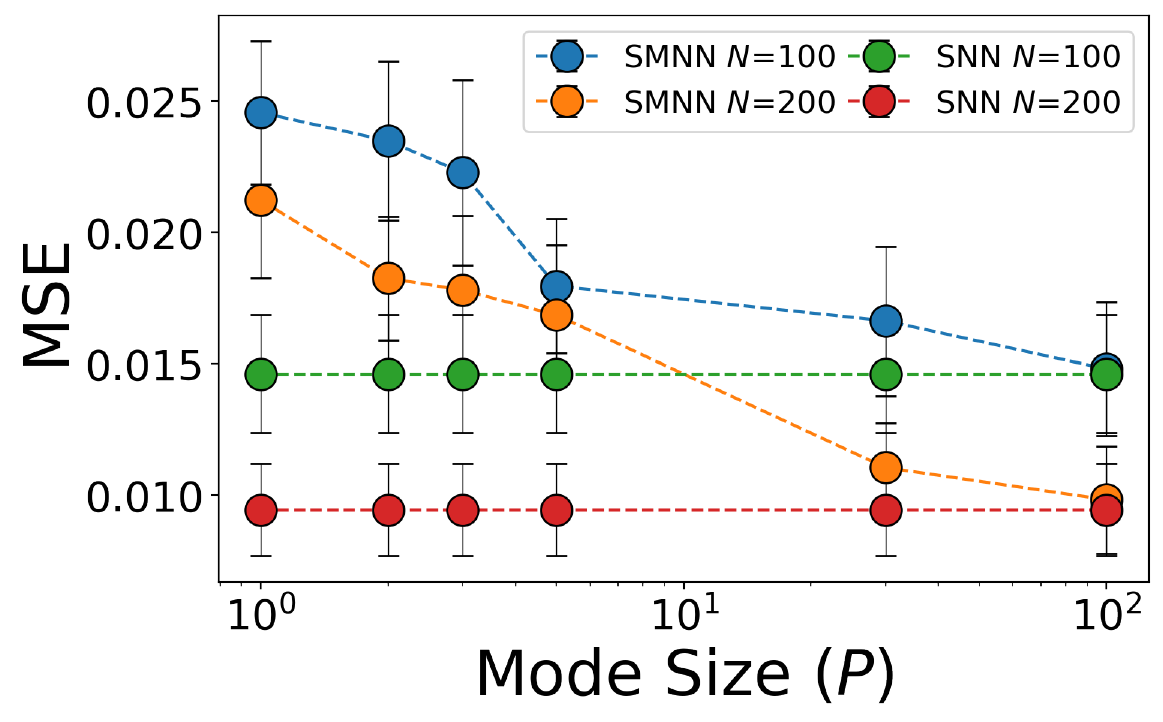}
}
\subfigure[]{
  \includegraphics[width = 0.3\linewidth]{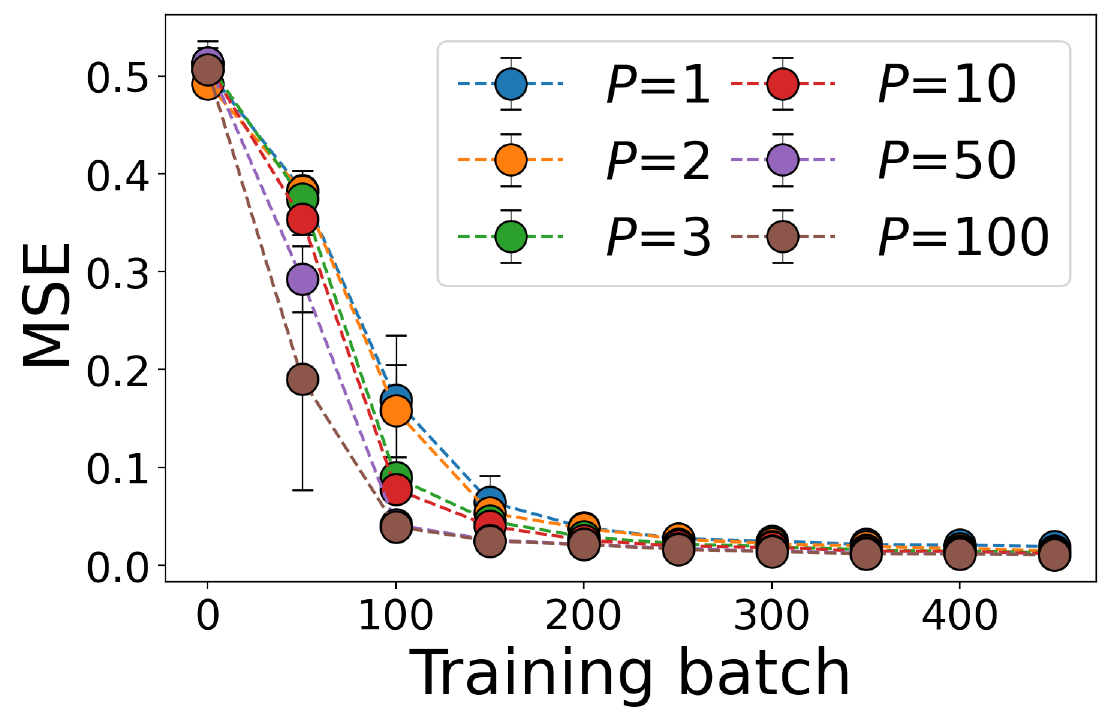}
}
\subfigure[]{
  \includegraphics[width = 0.3\linewidth]{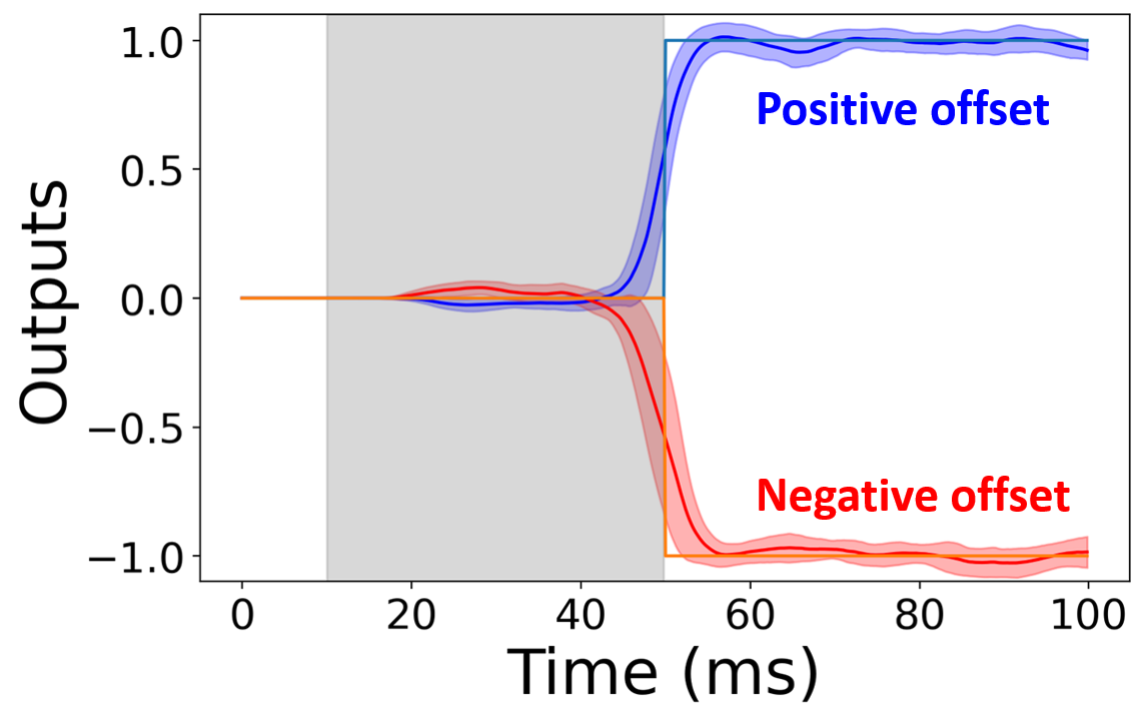}
}
\subfigure[]{
  \includegraphics[width = 0.45\linewidth]{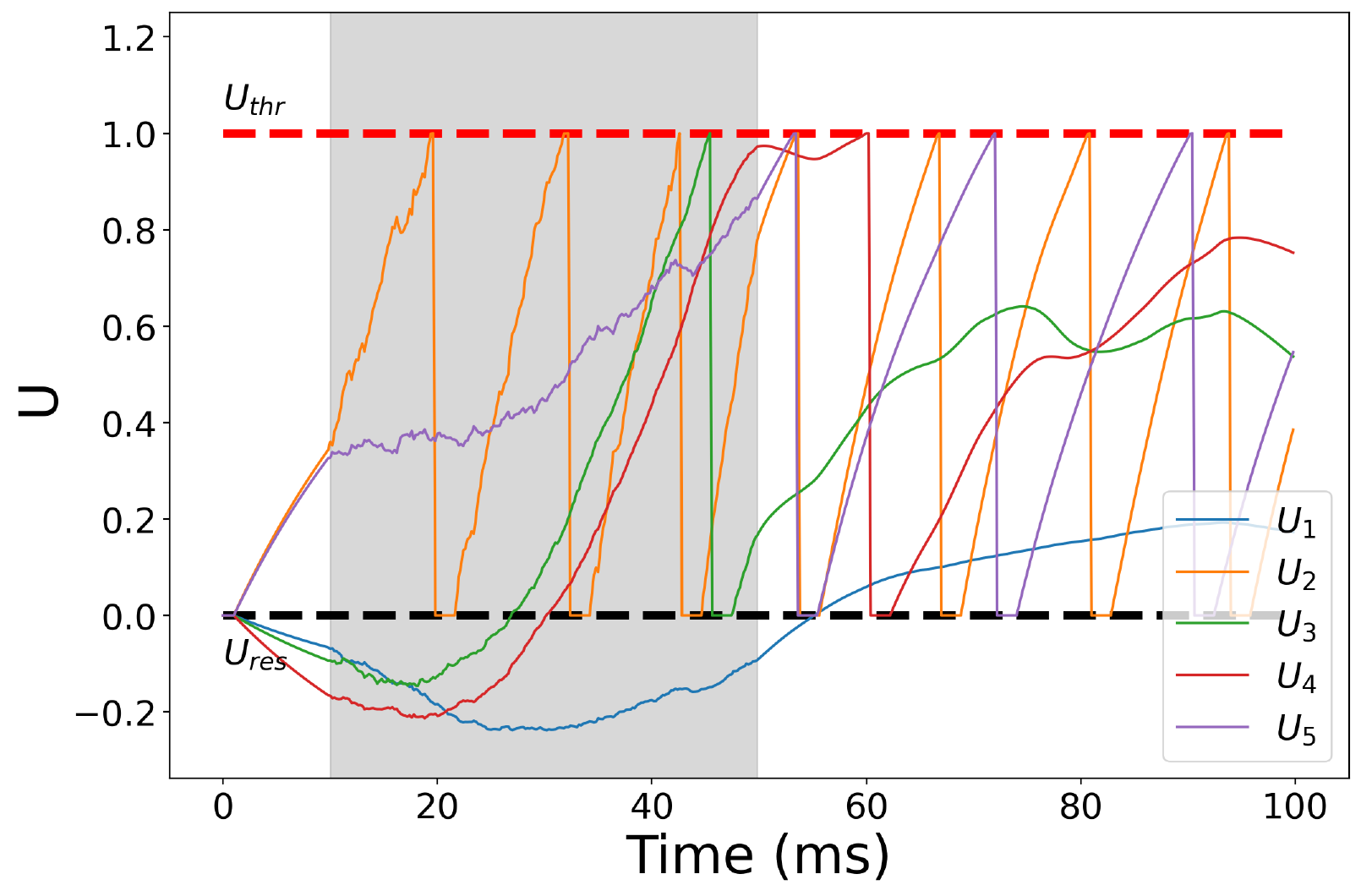}
}
\subfigure[]{
  \includegraphics[width = 0.45\linewidth]{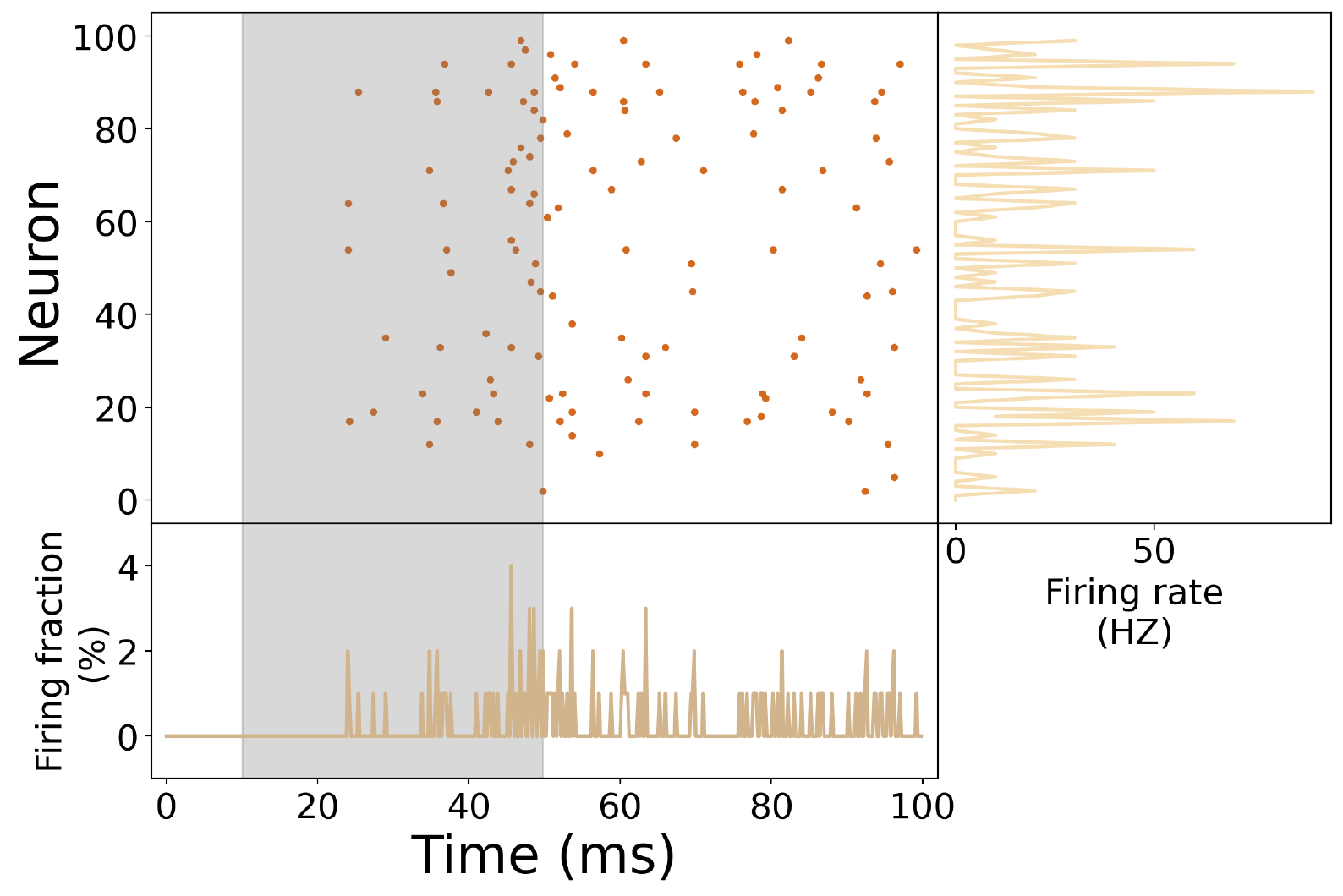}
}
  \caption{Learning performance of contextual integration task. Five independent runs are used to estimate the standard deviation. (a) Mean squared error (MSE) versus different mode sizes. Spiking model without the mode decomposition learning (this counterpart is called SNN) is compared. SMNN indicates the mode-based learning. (b) MSE as a function of training mini-batch for different mode sizes. The network size $N$=100. (c) Average output activity in response to test inputs for $(P,N) = (3,100)$. The shaded region indicates the stimulus period. Sensory inputs are only shown during the stimulus period, followed by a response period. Before the response period, the target output is always set to zero. The shaded region indicates the stimulus period. The fluctuation over $100$ random trials is also shown. Colored lines are two target outputs. (d) Membrane potential trace for five typical reservoir neurons in response to a random input.  (e) Spike raster of reservoir neurons with neuron firing rate (right) and population firing fraction (bottom). $(P,N) = (3,100)$ for both (d) and (e). }
  \label{ctx}
\end{figure}

As shown in Figure~\ref{ctx} (a), the test MSE [i.e., $\mathcal{L}^2$, see Eq.~\eqref{manteloss}] for the SMNN decreases with the mode size. In particular, a larger neural population is better. Even for a few modes like $P=3$, the predicted output already matches the target very well [see Figure~\ref{ctx} (c)]. The SNN without the mode decomposition achieves a lower MSE, but we emphasize that the accuracies behind the MSE within the shown range (from $0.01$ to $0.02$) do not look very different [see Figure~\ref{ctx} (c) for the case of $P=3$]. In this sense, compared to SNN, SMNN saves the training parameters. The training dynamics in Figure~\ref{ctx} (b) shows that a larger value of $P$ speeds up the learning. Figure~\ref{ctx} (c) confirms that our training protocol succeeds in reproducing the result of Monkey's experiments on the task of flexible selective integration of sensory inputs. We can even look at the dynamics profile of the membrane potential [Figure~\ref{ctx} (d)]. During the 
stimulus period, each neuron has its own time scale to encode the input signals, and at a network scale, we do not observe any regular patterns but an asynchronous pattern typically observed in prefrontal cortex. After the 
sensory input is turned off, the network is immediately required to make a decision, and we observe that the firing frequency of the neural
pool is elevated already a bit earlier than the moment of removing input, but a very low population firing rate is still maintained [Figure~\ref{ctx} (e)]. This observation is a reflection of neural dynamics on the neural manifolds underlying the perceptual decision making, which we shall detail below.

\subsection{Power law for connectivity importance scores}
\begin{figure}
\centering
  \subfigure[]{
  \includegraphics[width = 0.42\linewidth]{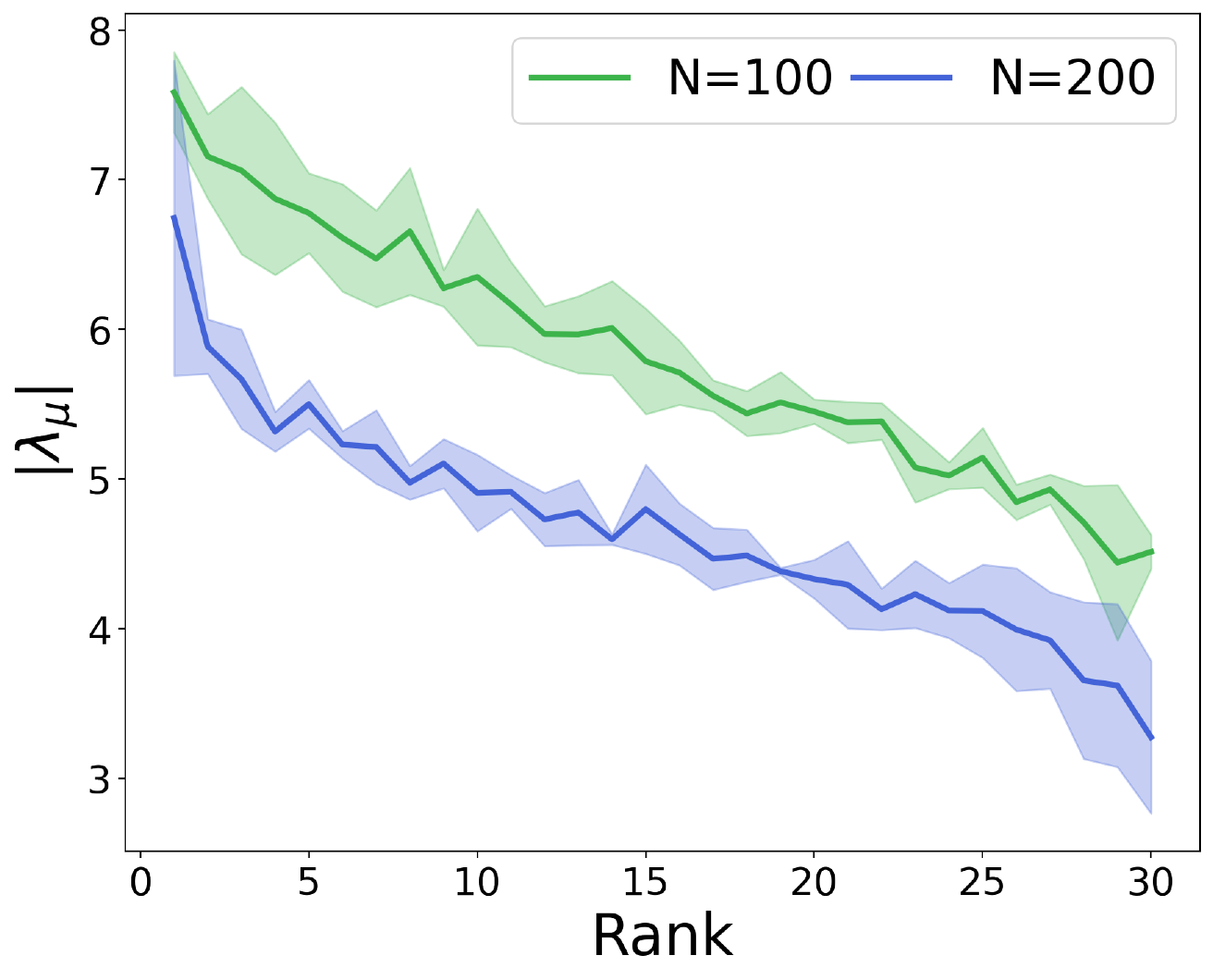}
}
\subfigure[]{
  \includegraphics[width = 0.45\linewidth]{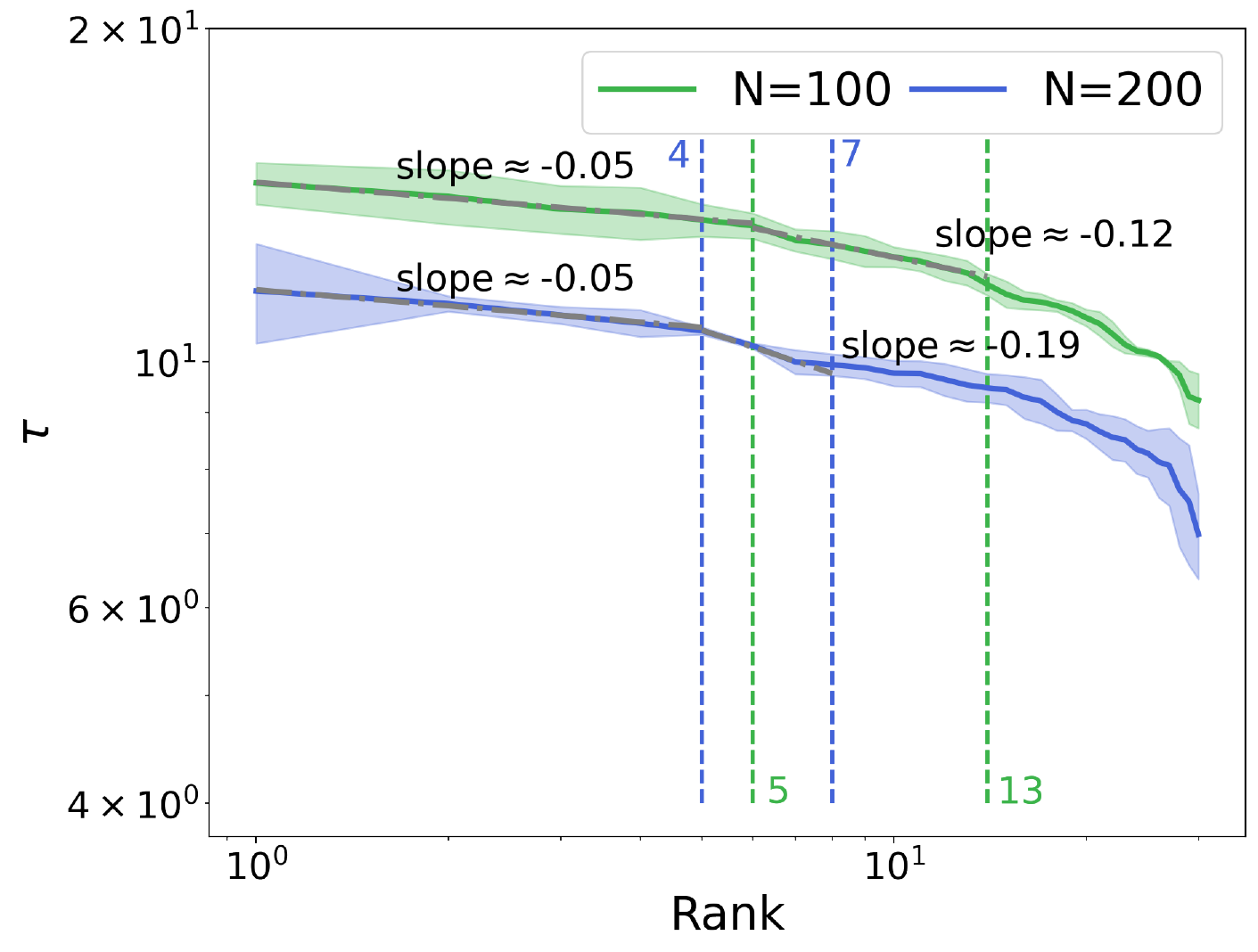}
  }
  \subfigure[]{
  \includegraphics[width = 0.42\linewidth]{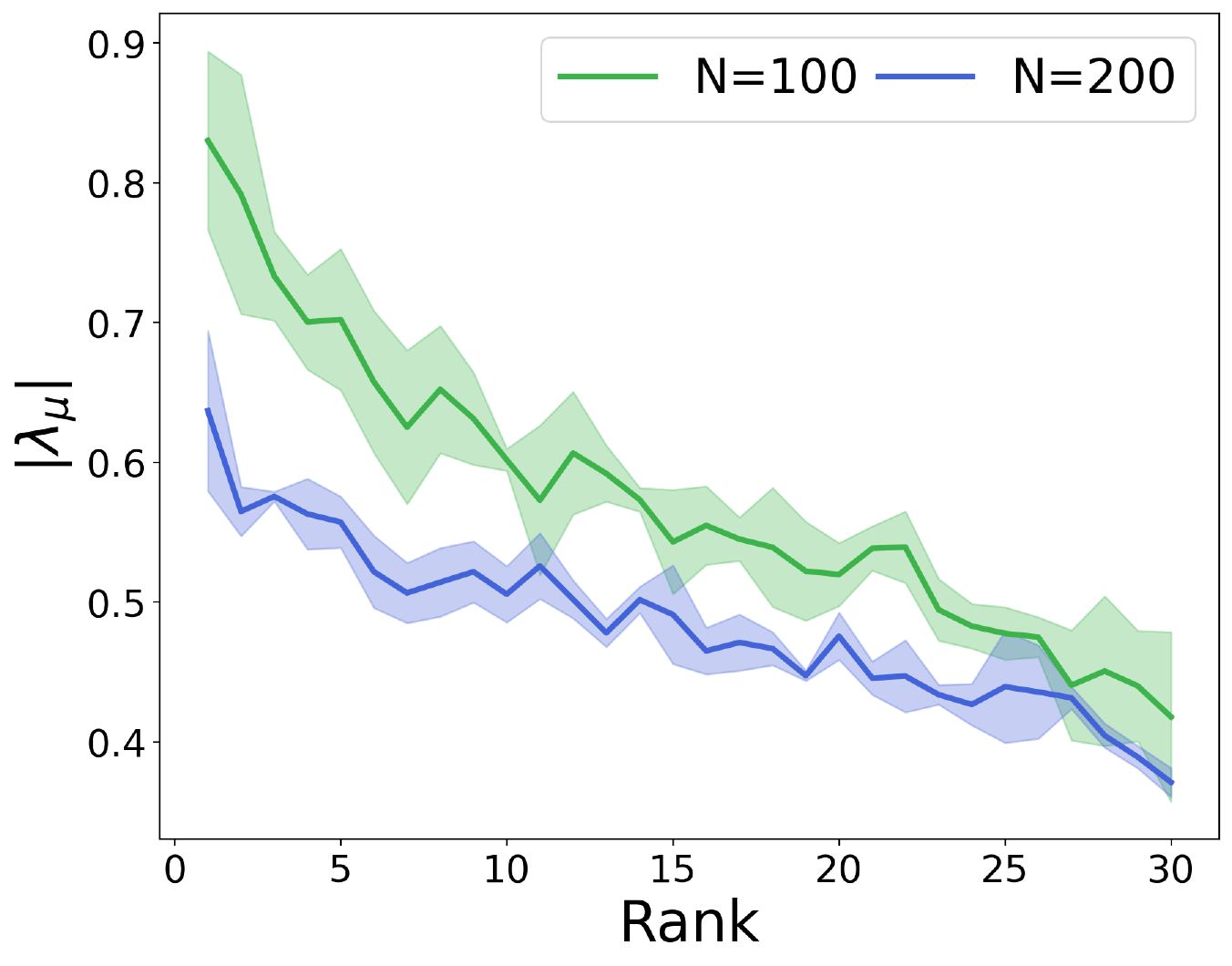}
}
\subfigure[]{
  \includegraphics[width = 0.45\linewidth]{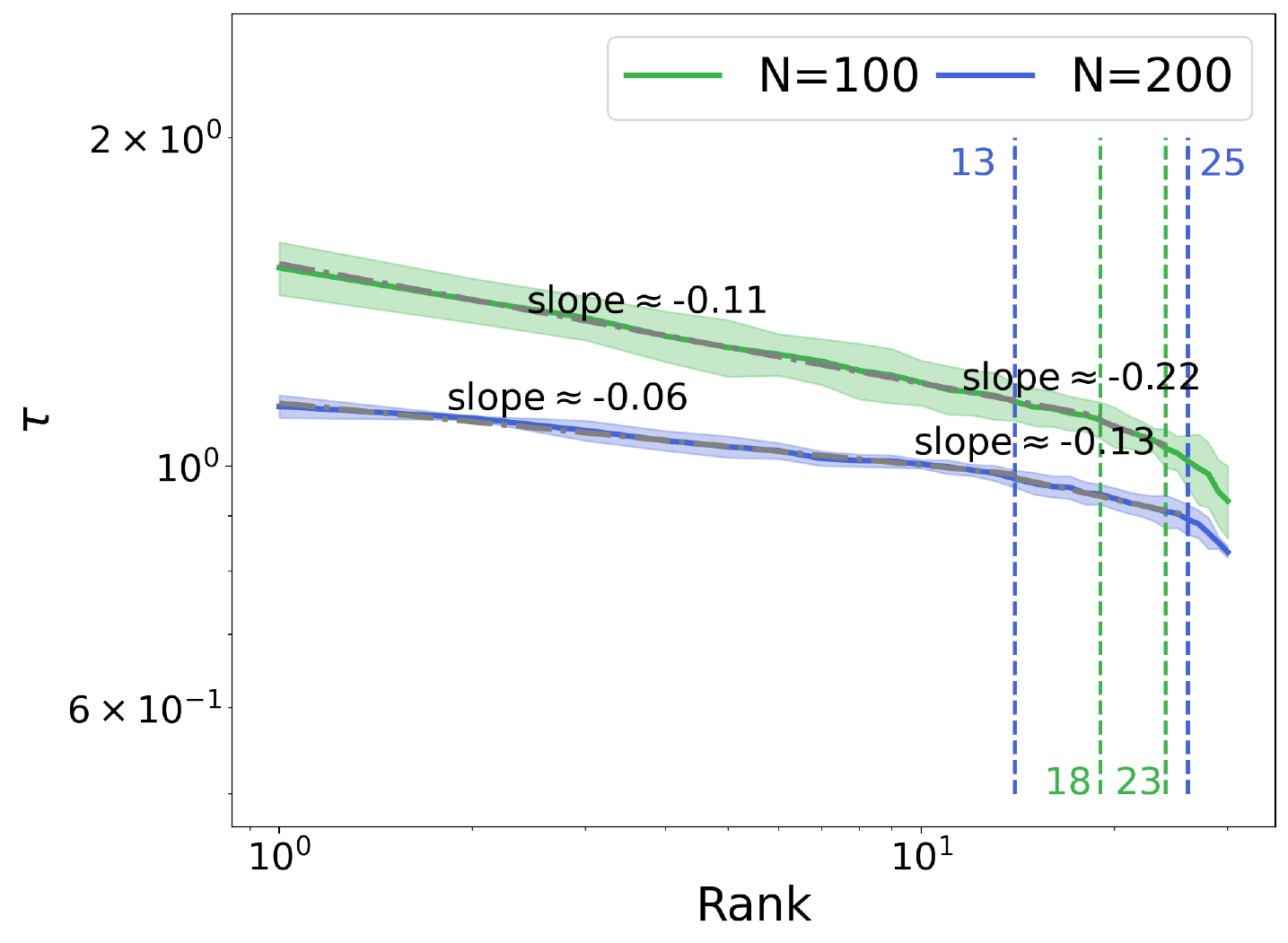}
}
  \caption{ Connectivity importance $|\lambda_\mu|$ versus rank (in descending order). Both MNIST classification and contextual integration task are considered, and the simulation conditions are the same with that in Figure~\ref{digit} for MNIST and that in Figure~\ref{ctx} for context dependent computation. We fix $P=30$.  Three independent runs are used to estimate the standard deviation. We also define a more precise measure $\tau_\mu=\chi\Vert\bm{\xi}^{\rm in}_\mu\Vert_2 + \chi\Vert\bm{\xi}^{\rm out}_\mu\Vert_2 + |\lambda_\mu|$, where $\chi=\sum_\mu |\lambda_\mu|/\sum_\mu(\Vert\bm{\xi}^{\rm in}_\mu\Vert_2+\Vert\bm{\xi}^{\rm out}_\mu\Vert_2)$. (a,b) MNIST classification. (c,d) Contextual integration task. There appears the piecewise power law behavior for the $\tau$-measure in the log-log plot (b,d). The colored vertical dashed lines mark the fitting ranges for different network sizes.}
  \label{pw}
\end{figure}

We next find whether some modes are more important than others. To make comparable the magnitudes of the pattern and importance scores, we rank the modes according to the following measure~\cite{Li-2023}
\begin{equation}
\tau_\mu=\chi||\bm{\xi}^{\rm in}_\mu||_2 + \chi||\bm{\xi}^{\rm out}_\mu||_2 + |\lambda_\mu|,   
\end{equation}
where $\bm{\xi}_\mu^{\rm{in}/\rm{out}}\in\mathbb{R}^N$ ($\mu$-th column of $\bm{\xi}^{\rm{in}/\rm{out}}$), and $\chi=\sum_\mu |\lambda_\mu|/\sum_\mu(\Vert\bm{\xi}^{\rm in}_\mu\Vert_2+\Vert\bm{\xi}^{\rm out}_\mu\Vert_2)$. We observe a piecewise power law for the $\tau$ measure (Figure~\ref{pw}), implying that the original state space of neural activity is actually low-dimensional, and can be projected to a low dimensional mode space, with $P_{\rm dom}$ (taking a small value compared to $N$ in Figure~\ref{pw}) dominant coordinate axes on which the $\tau$-measure vary mildly. For visualization (see next subsection), we can only take $P\le3$, whose results can be compared with experimental intuition. On one hand, the task information is coded hierarchically into the mode space, and on the other hand, this observation supports that a fast training of spiking networks is possible by focusing on leading modes explaining the network macroscopic behavior.

\subsection{Reduced dynamics in the mode space}
Here, we will show how the dynamics of network activity based on our SMNN learning can be projected to a low-dimensional counterpart. We follow the previous works of low-rank recurrent neural networks~\cite{Ostojic-2018,Ostojic-2023b}. To construct a subspace spanned by orthogonal bases, we first split each column of $\bm{W}^{\rm in}$ ($\in\mathbb{R}^{N\times N_{\rm in}}$) into the parts parallel and orthogonal to the input mode $\bm{\xi}^{\rm in}$ (or the output mode),
\begin{equation}
\bm{W}^{\rm in}_s =\sum_\mu \alpha_\mu \bm{\xi}^{\rm in}_\mu+ \beta_s\bm{W}_{\bot}^s,
\end{equation}
where the $\mu$-th column of $\bm{\xi}^{\rm{in}}$ is denoted by $\bm{\xi}_\mu^{\rm{in}}\in\mathbb{R}^N$, $s$ indicates which component of input signals, $\alpha_\mu$ and $\beta_s$ are coefficients for the linear combination. We assume that the bases are orthogonal; otherwise, one can use Gram-Schmidt procedure to obtain the orthogonal bases.

 The filtered spike trains $\ra(t)$ can then be written as a form of linear combination,
\begin{equation}
\ra(t) = \sum_\mu\kappa_\mu(t)\bm{\xi}^{\rm in}_\mu +\sum_s\nu_s(t) \bm{W}_{\bot}^s,
\end{equation}
where the coefficients are given respectively by
\begin{equation}
\begin{aligned}
\kappa_\mu(t) &= \frac{(\bm{\xi}^{\rm in}_\mu)^\top\ra(t)}{(\bm{\xi}^{\rm in}_\mu)^\top\bm{\xi}^{\rm in}_\mu},\\
\nu_s(t) &= \frac{(\bm{W}_{\bot}^s)^\top\ra(t)}{(\bm{W}_{\bot}^s)^\top\bm{W}_{\bot}^s}. 
\end{aligned}
\end{equation}
The coefficients $\{\kappa_\mu(t)\}$ forms a low-dimensional latent dynamics of the computational task, for which we display the results for the two tasks in the following section.

\subsection{Projection in the mode space}
\begin{figure}
\centering
\subfigure[]{
  \includegraphics[width = 0.45\linewidth]{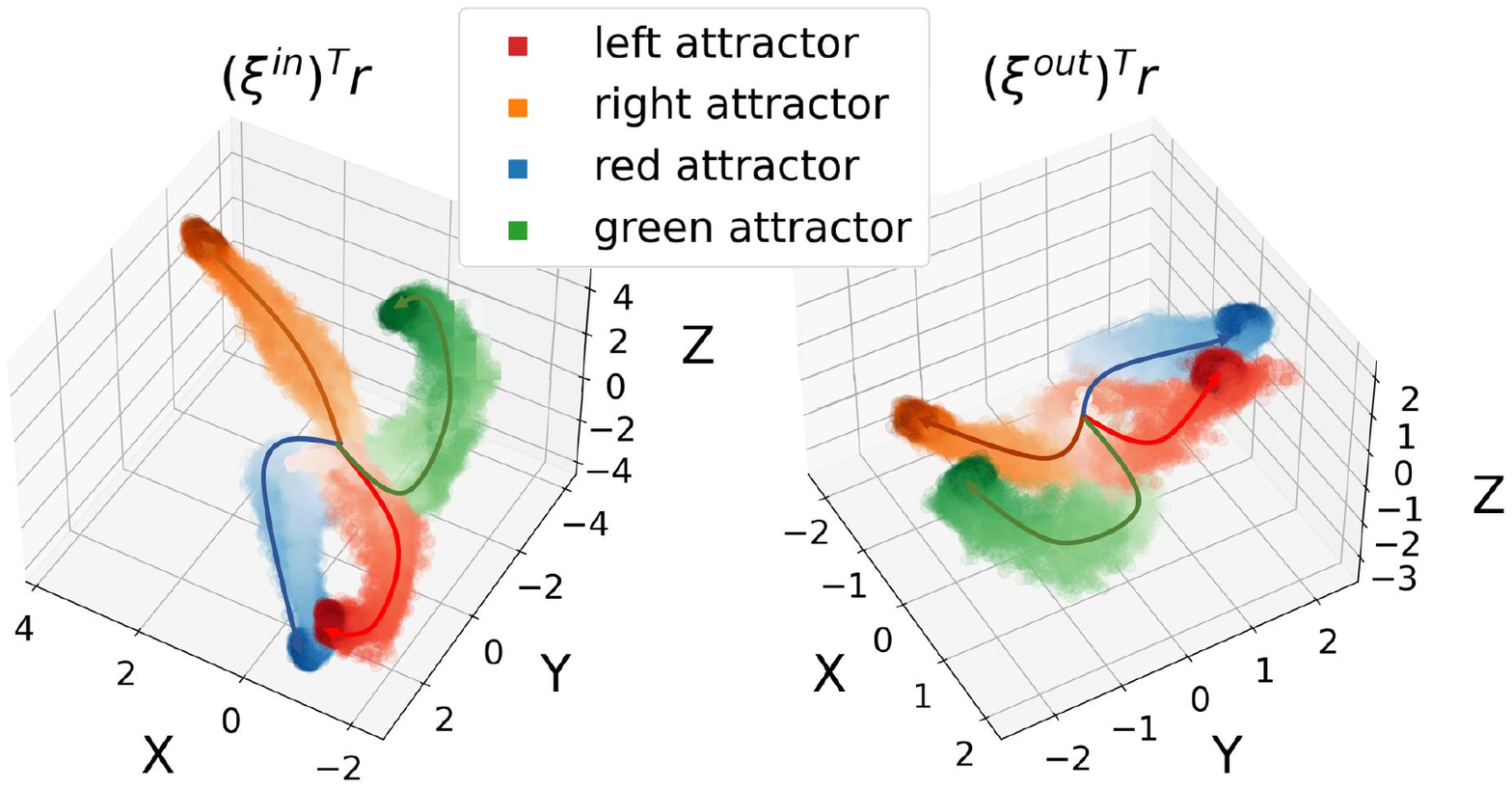}
}
\subfigure[]{
  \includegraphics[width = 0.45\linewidth]{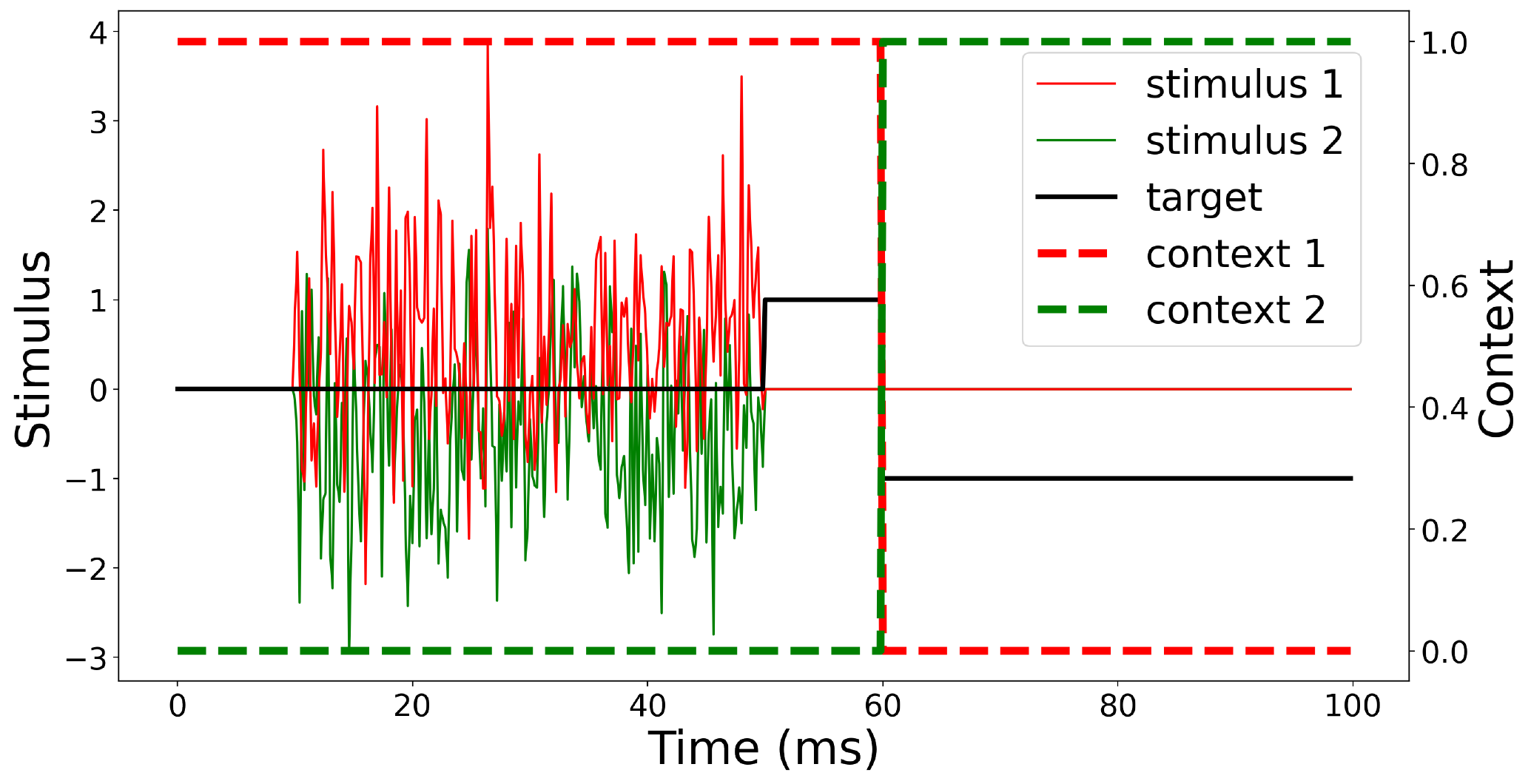}
}
\subfigure[]{
  \includegraphics[width = 0.55\linewidth]{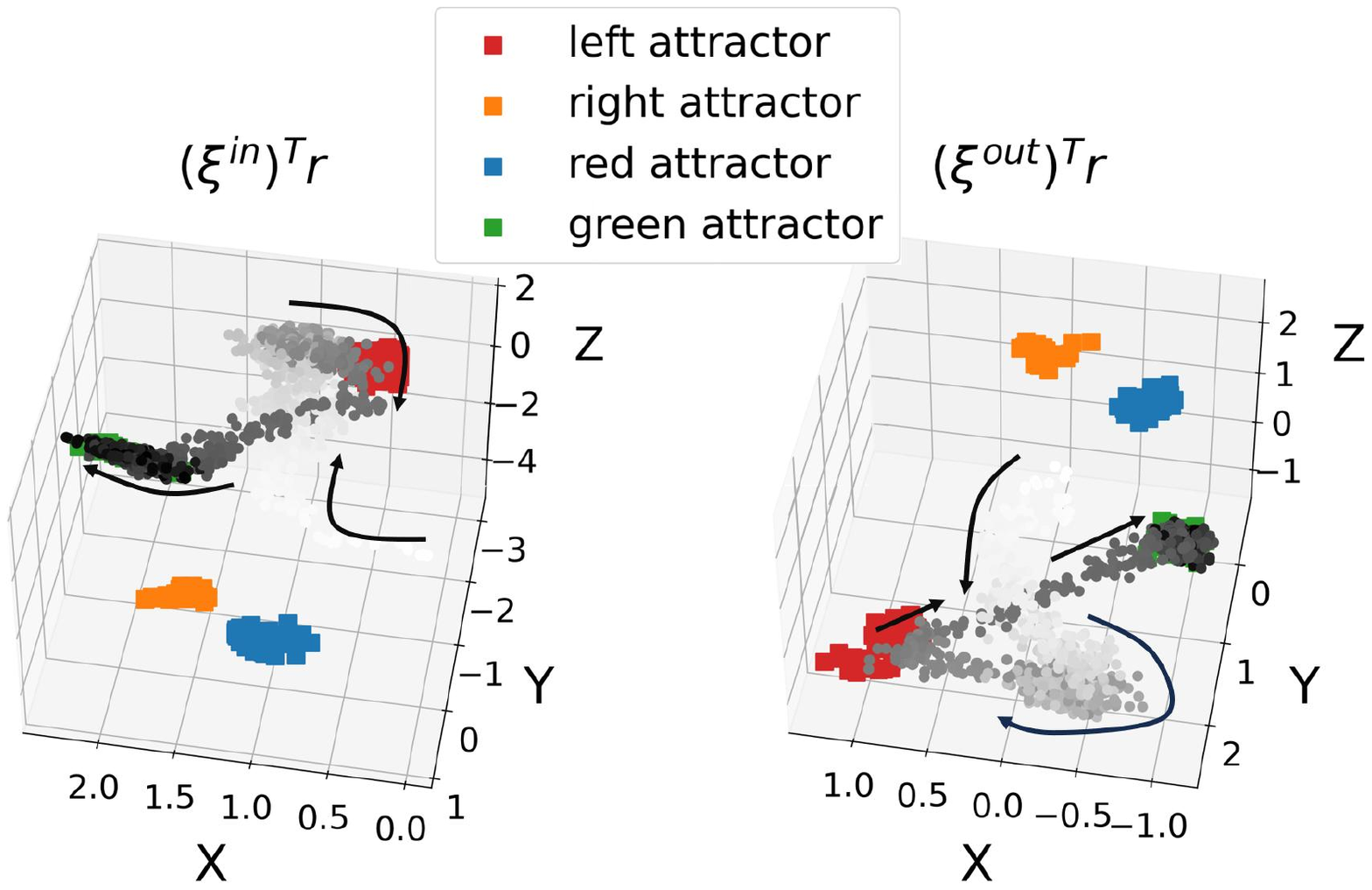}
}
\subfigure[]{
  \includegraphics[width = 0.4\linewidth]{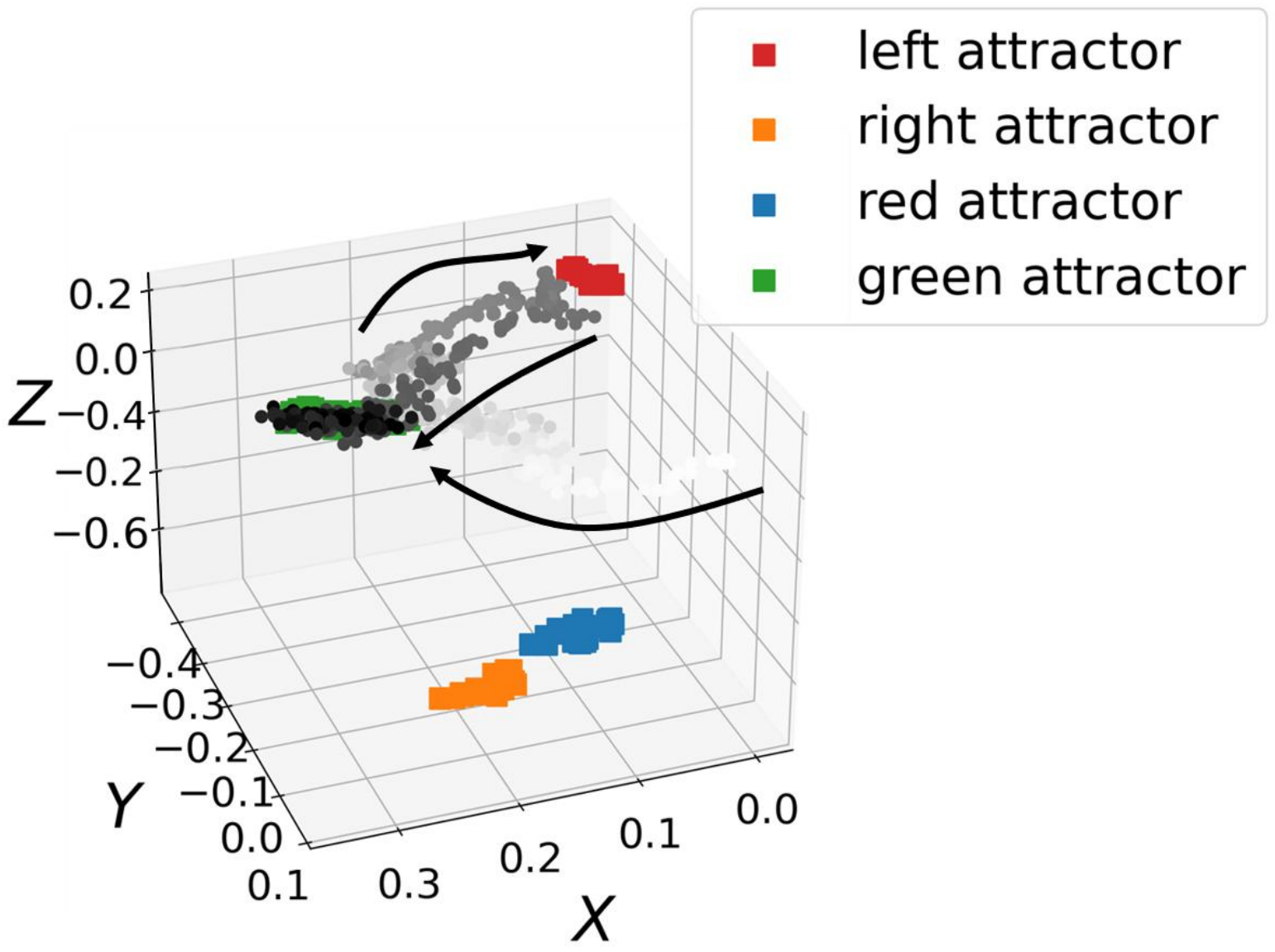}
}

  \caption{The filtered spike train of  the hidden layer projected to the mode space for the contextual integration task. The mode size $P=3$, and the network size $N = 100$.  (a) Projection into the input mode space.  Three hundred randomly generated trials were used. Different colors encode different offset signs and contextual cues. The color gets darker with time in the dynamics trajectory. (b) Context switching experiment. At $t = 30\,\rm{ms}$, the previous
   contextual cue is shifted to the other one. The left-y axis encodes the input signals, while the right-y axis encodes the contextual information.
     (c) Activity projection for the context switching experiment in (b). (d) Projection coefficient in the input mode space for the context switching experiment in (b).}
     \label{proj}
\end{figure}

To study the neural manifold underlying the perceptual decision-making, we project the high dimensional neural activity onto the mode space. If the small number of modes is sufficient to capture the performance, we can visualize the manifold by projecting the neural activity onto the mode space, using the input or output mode vector as the basis.  As expected, after training, four separated attractors are formed in the mode
space, either in the input mode space or in the output mode space [Figure~\ref{proj} (a)]. The test dynamics would flow to the corresponding attractor depending on the context of the task.

\begin{figure}
\centering
  \includegraphics[width = 0.8\linewidth]{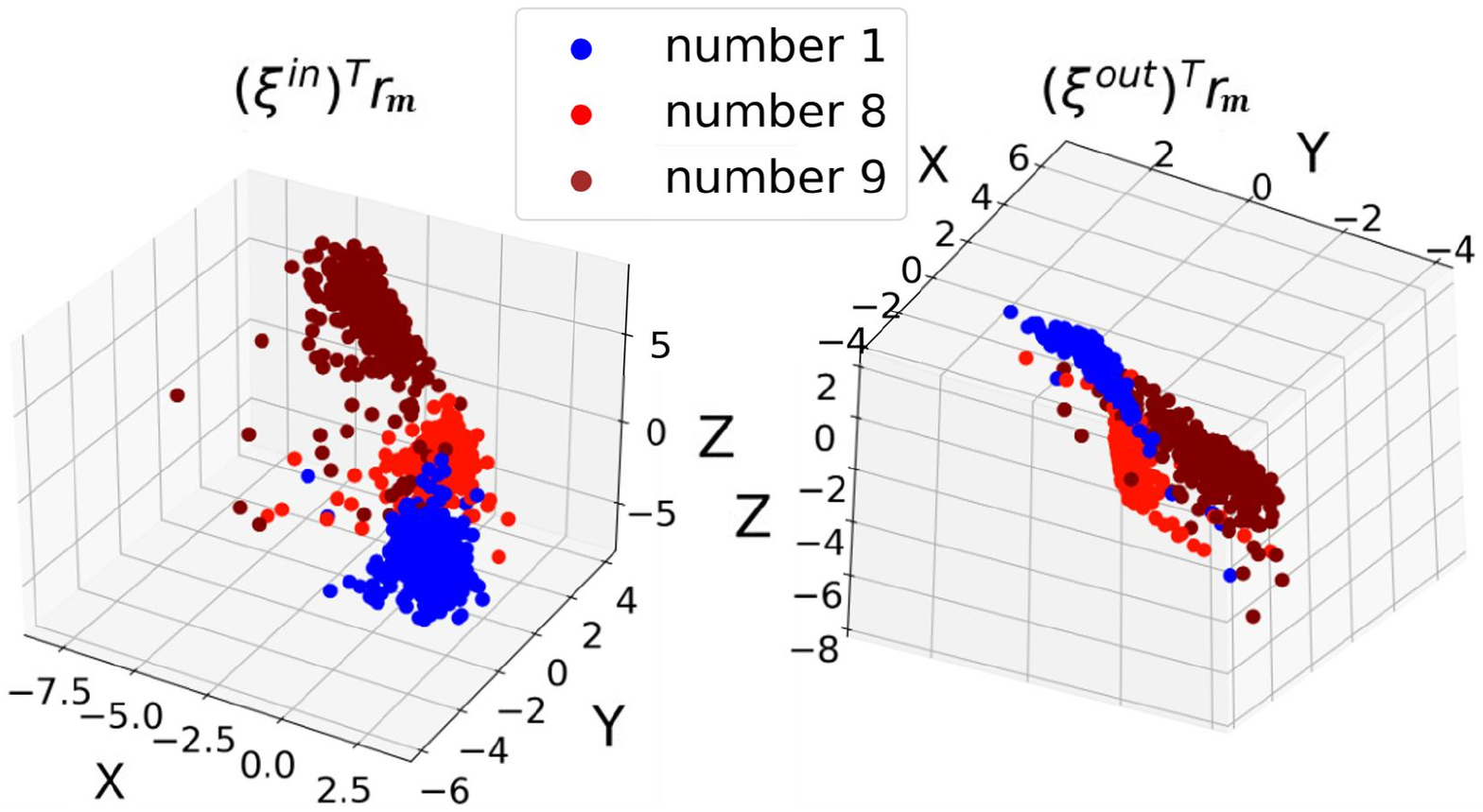}
  \caption{ Projection of the network activity at the moment of maximum firing rates in the MNIST classification task. Three labels of digits are considered. Other settings are the same as in Fig.~\ref{digit}. }
  \label{digproj}
\end{figure}

We next consider a context switching experiment, and look at how the dynamics is changed in the low dimensional intrinsic space.
The experimental protocol is shown in Figure~\ref{proj} (b). At $t = 30\, \rm{ms}$, the context is switched to the other one, and the network should carry out
computation according to the new context, e.g., making a correct response to the input signal.
Before the context is switched, the context cued input signals have a positive offset. Correspondingly, the neural activity trajectory in the mode space evolves to the left attractor [Figure~\ref{proj} (c)]. Once the context is switched, the context cues another input signal that has a negative offset. The neural trajectory is then directed to the neighboring green attractor. Therefore, the neural dynamics can be guided by the contextual cue, mimicking what occurs in the prefrontal cortex of Monkeys that  perform the contextual integration task~\cite{Mante-2013}. We plot the dynamics of $\boldsymbol{\kappa}$ for the context switching experiment [Figure~\ref{proj}(d)], which reveals the qualitatively same behavior as observed in Figure~\ref{proj} (c). A similar attractor picture is also found in the MNIST classification task (Figure~\ref{digproj}).

\subsection{Spiking variability measured by Fano factor}
The spiking activity is commonly irregular, displaying strong variability even when the neural population is shown identical stimuli. One can thus count the spiking 
events in a time window of duration $T$ for one neuron, and then repeat the experiment over many trials and finally measure the mean and variance of the counts. The spiking variability is measured by
 the Fano factor calculated as~\cite{ND-2014}
\begin{equation}
   F(T)=\frac{\sigma_{N}^{2}(T)}{\mu_{N}(T)},
\end{equation}  
where $\mu_N$ is the mean value of spike counts for one neuron in a neural population of $N$ neurons, while $\sigma^2_N$ denotes the corresponding variance of spike counts.
The Fano factor is exactly one for a Poisson spiking process, irrespective of $T$.

\begin{figure}
\centering
  \subfigure[]{
  \includegraphics[width = 0.45\linewidth]{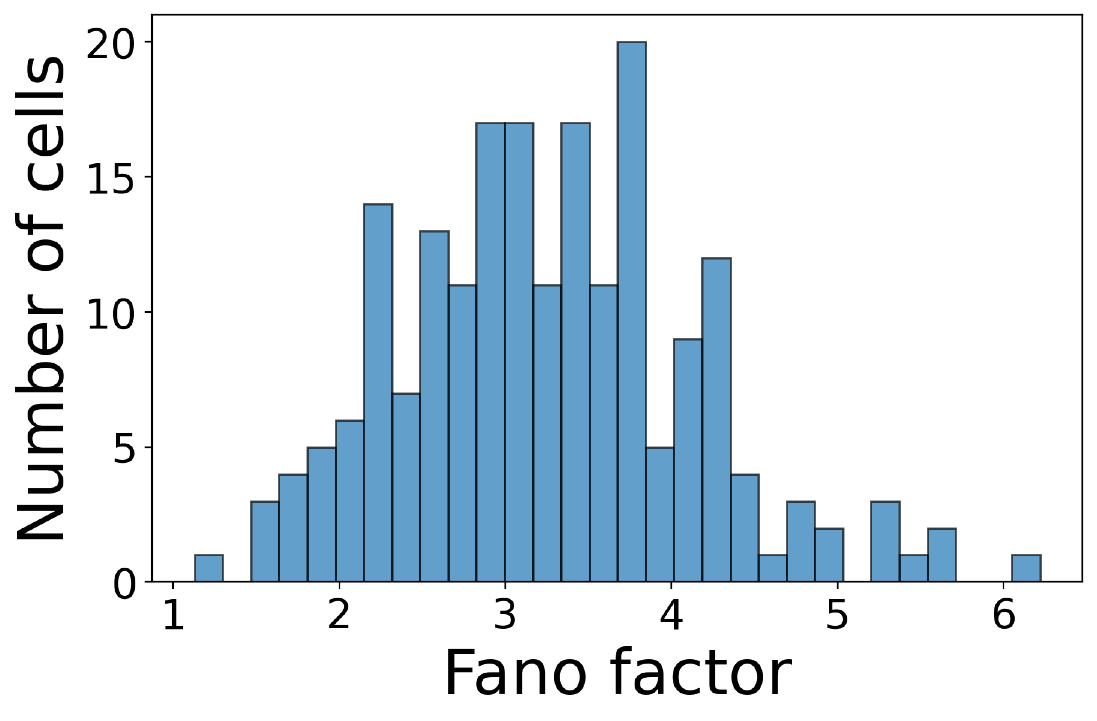}
}
\subfigure[]{
  \includegraphics[width = 0.45\linewidth]{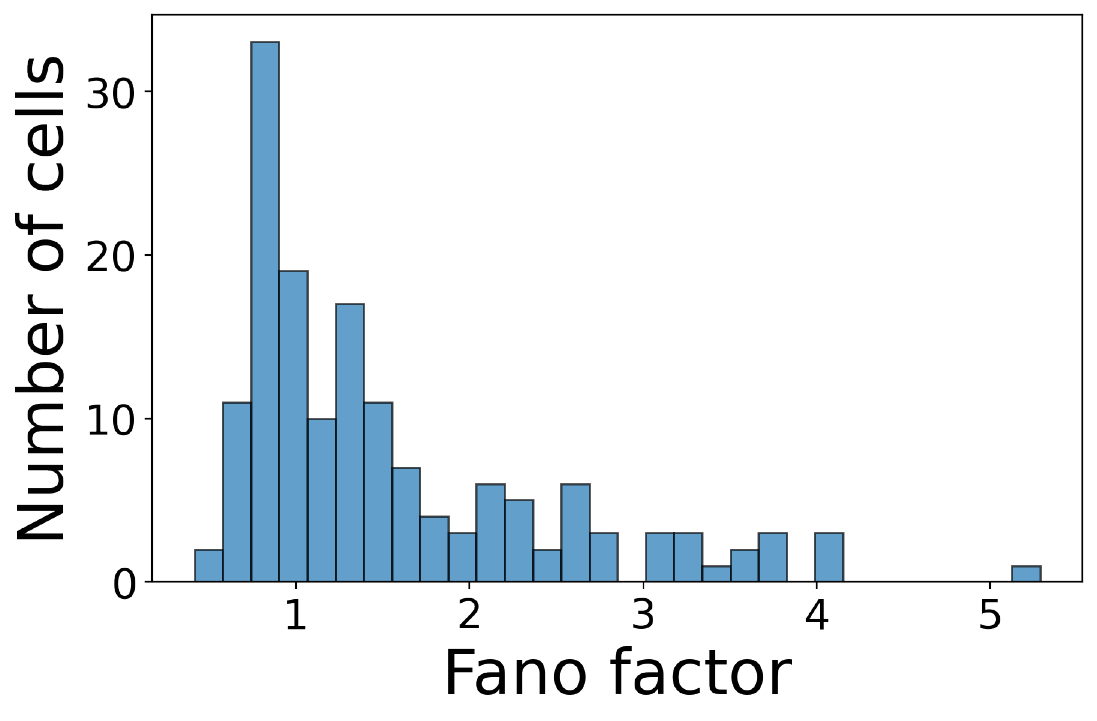}
}
  \caption{ Fano factor statistics across neurons.
(a) MNIST classification. (b) Contextual integration task. $(P,N)=(10,200)$.  }
  \label{fano}
\end{figure}

We plot the Fano factor distribution across neurons in Fig.~\ref{fano}. The Fano factor is estimated over the entire experimental window.
For the MNIST classification task, the Fano factor is mainly distributed from $2$ to $4$ [Figure~\ref{fano} (a)], indicating highly unreliable firing patterns, which corresponds to that observed in Figure~\ref{digit} (e). For an input image, some neurons fire at high frequencies, while others remain silent. This non-Poissonian phenomenon has also been observed in cortical areas~\cite{Pos-2009}. For the context-integration task, the 
Fano factor is surprisingly concentrated around one [Figure~\ref{fano} (b)], except for a long tail at large values of Fano factor. This model observation highlights the potential of our method in studying the alignment between the simulation and neural experiments of context dependent cognitive decision making. 

\subsection{Ablation studies on the refractory, activation function and network robustness}
\begin{figure}
\centering
  \subfigure[]{
  \includegraphics[width = 0.45\linewidth]{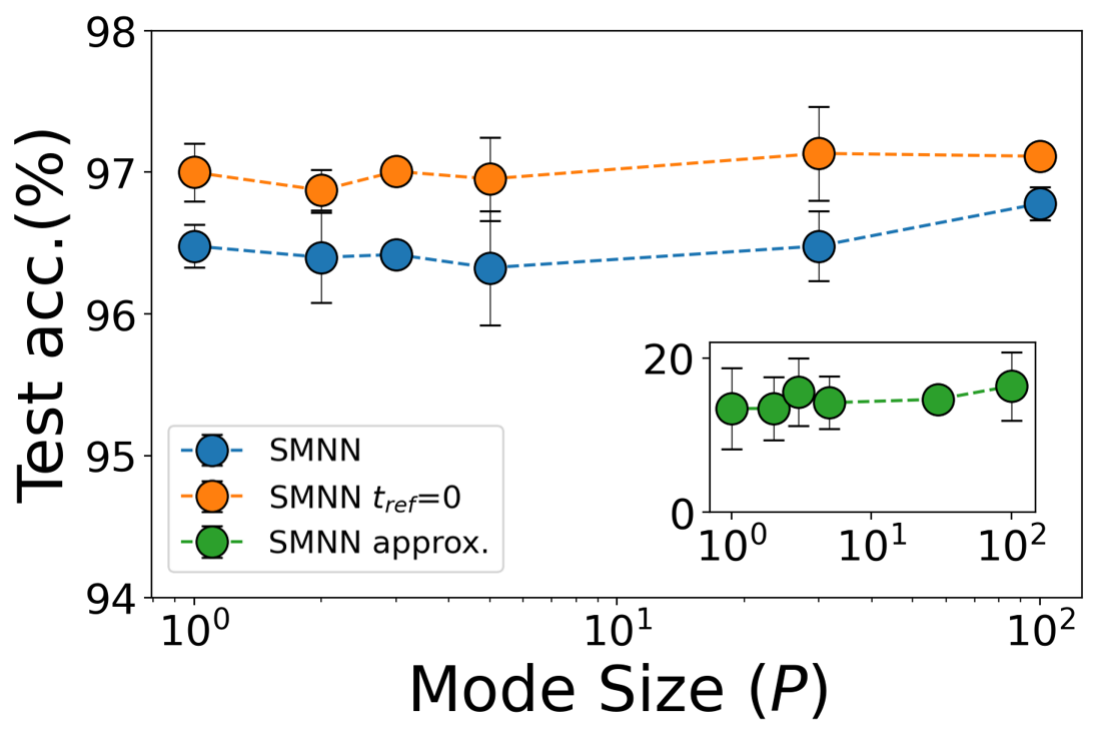}
}
\subfigure[]{
  \includegraphics[width = 0.45\linewidth]{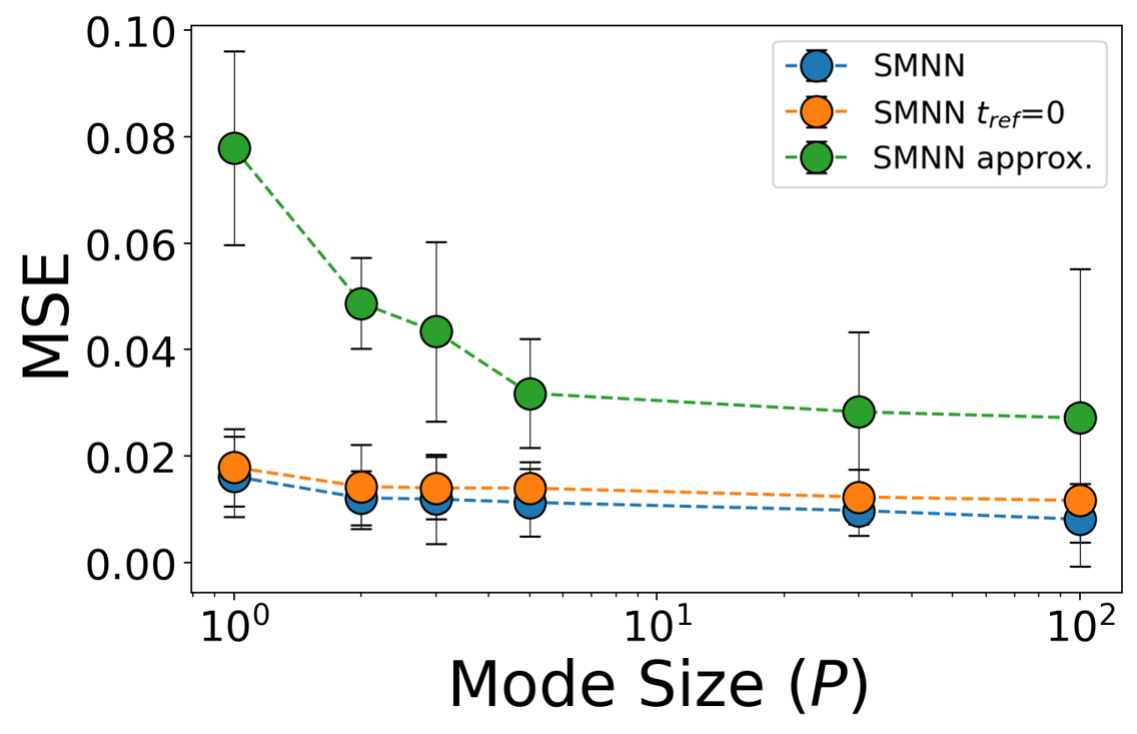}
}
  \caption{ Ablation experiments by setting the refractory period $t_{\rm ref}=0$ or replacing the step function by a smooth approximate function. The network size $N=100$, and the standard deviation is estimated using three independent experiments. (a) Test accuracy for the MNIST classification task. (b) Test mean squared error for the contextual integration task.
}
  \label{ab}
\end{figure}

To study effects of the refractory period and smoothness of activation function on the network performance, we further design ablation experiments, where we consider the first scenario of setting $t_{\rm ref}$ zero, and the second scenario of adopting a smooth current-transfer function $\phi(x)$ instead of the step function as follows,
\begin{equation}
    \phi(x)=\half [1+\tanh(\beta x)],
\end{equation}
where we set $\beta=20$ to mimic the step function. Note that in this control experiment, we use this transfer function both in forward and backward passes during the training.
 As shown in Figure~\ref{ab} (a) for the MNIST task, the absence of the refractory period allows the network to respond more correctly to external stimuli. Performance without the refractory period is slightly better than with the refractory period, but as the mode number increases, the gap shrinks.  For the contextual integration task, an appropriate refractory period is able to optimize network performance [Figure \ref{ab} (b)], consistent with previous results in standard SNNs~\cite{Kim-2019}. Finally, we remark that networks that use a smooth transfer function instead of the step function perform much worse than spiking networks using a sharp step function (neurons can thus fire or be silent), indicating the importance of spiking event.

\begin{figure}
\centering
  \subfigure[]{
  \includegraphics[width = 0.45\linewidth]{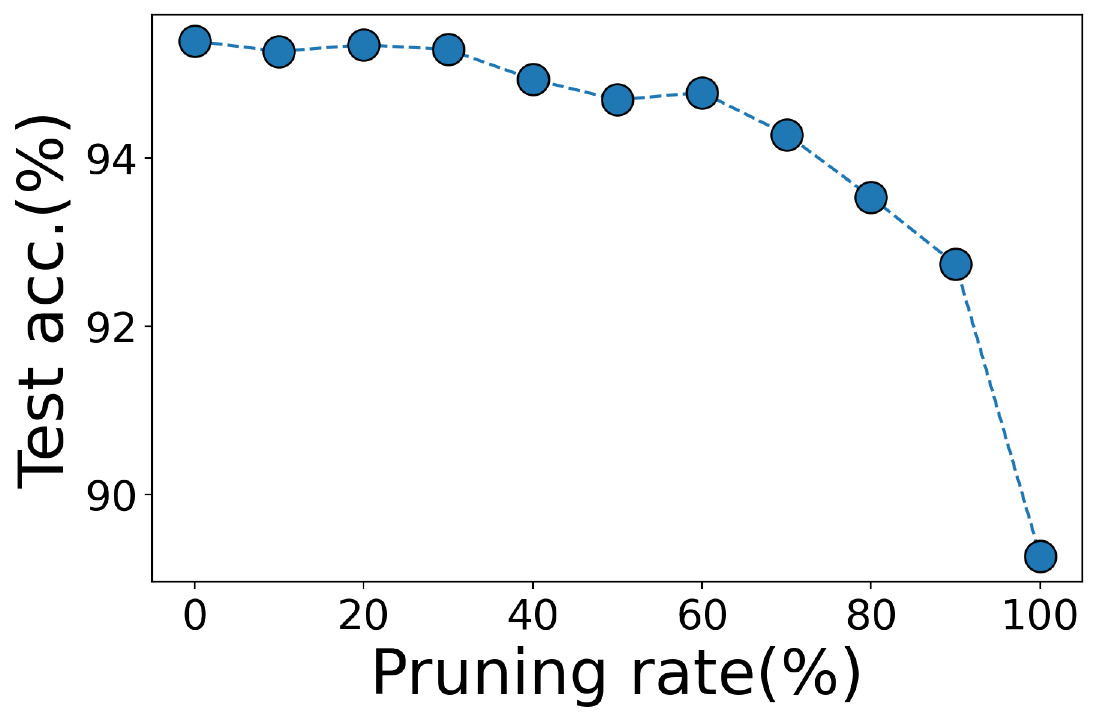}
}
\subfigure[]{
  \includegraphics[width = 0.45\linewidth]{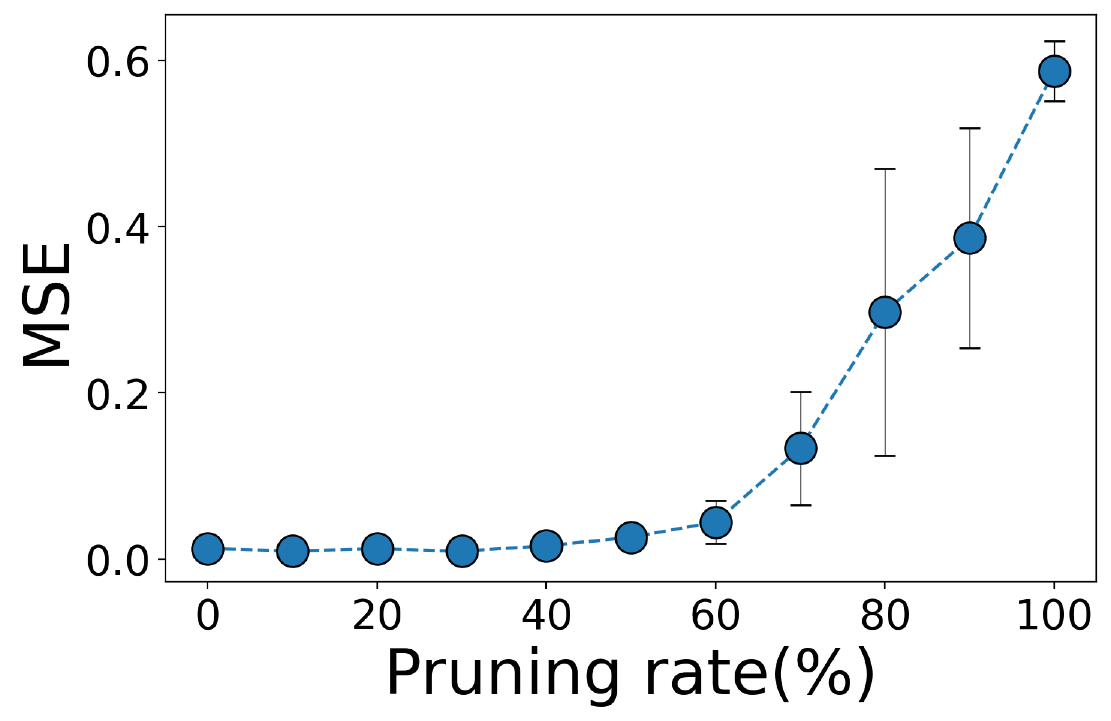}
}
  \caption{ Test performance in pruning experiments where the weights are set to zero if its absolute value falls below a threshold.  (a) Test accuracy for the MNIST classification task. (b) Test mean square error for the contextual integration task.  $(P,N)=(3,100)$.
}
  \label{prune}
\end{figure}

We next consider the network robustness in a weight-pruning experiment.
More precisely, for a well-trained connectivity matrix $W^{\rm rec}_{ij} =\sum_\mu \lambda_\mu\xi^{\rm in}_{i\mu}\xi^{\rm out}_{j\mu}$, we set some elements where their absolute values $|W^{\rm rec}_{ij}|<\theta$ to zero, where $\theta$ is an adjustable pruning threshold. This operation leads to a diluted connectivity matrix $\hat{\bm{W}}^{\rm rec}$:
\begin{equation}
    \hat{\bm{W}}^{\rm rec}=\bm{W}^{\rm rec}\odot \bm{M},
\end{equation}
where $\odot$ is an element-wise multiplication, and  the $\theta$-tuned mask matrix reads
\begin{equation}
    M_{ij}=\left\{
\begin{aligned}
0,\ & \ |W^{\rm rec}_{ij}|<\theta;\\
1,\ & \ |W^{\rm rec}_{ij}|\geq\theta.
\end{aligned}
\right.
\end{equation}
Increasing the threshold $\theta$ increases the pruning rate, making the original network sparser. 
As shown in Figure~\ref{prune},  for the MNIST classification task and the contextual integration task, when the pruning rate is below 60\%, there is no significant change in the network's test accuracy (or mean squared error). However, when the pruning rate is above 60\%, the network's test accuracy (or mean squared error) sharply decreases (or increases) with the pruning rate, indicating that a strongly diluted network sacrifices the network performance. But this raises another interesting open question whether a diluted network can be learned while maintaining a similar performance to the full network, rather than pruning a well-trained weight matrix of full rank. Training a sparse mode-based spiking network seems challenging as the sparsity of the connectivity increases~\cite{Brian-2023}.  Interestingly, even with complete pruning of connection weights, the test accuracy for the MNIST classification task remains close to 89\%. This might be explained by the specific transformation layer which already contains enough information about the digit label and therefore a rate-maximum-over-time readout is accurate.

\begin{figure}
\centering
  \subfigure[]{
  \includegraphics[width = 0.45\linewidth]{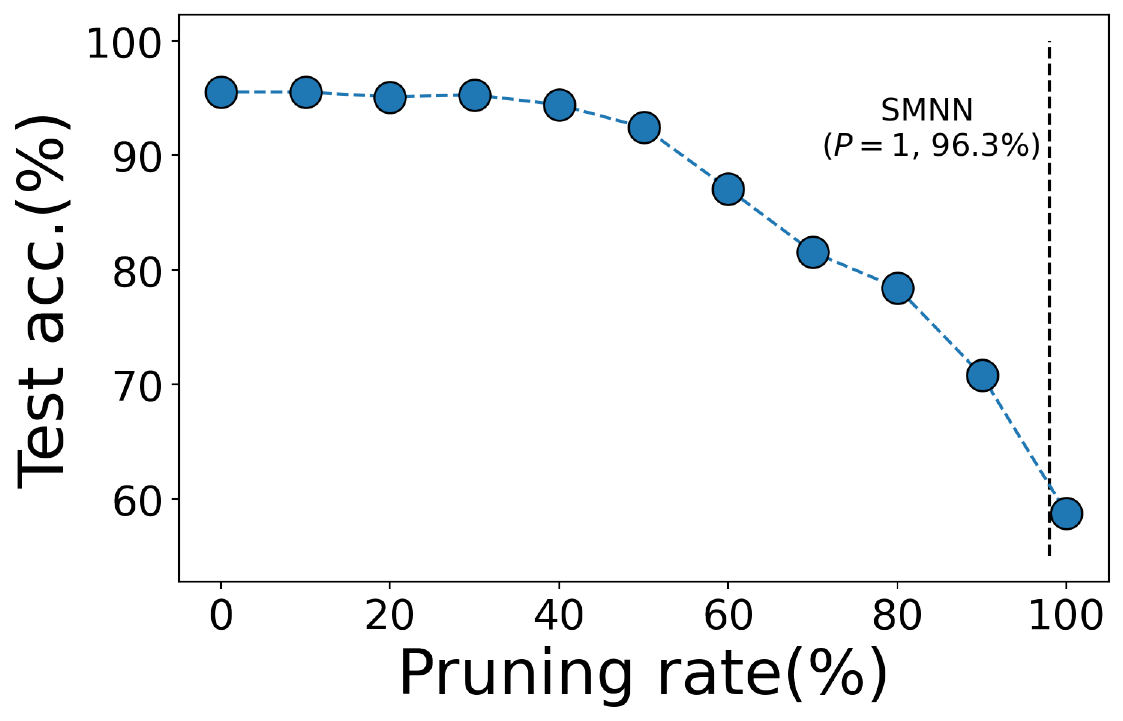}
}
\subfigure[]{
  \includegraphics[width = 0.45\linewidth]{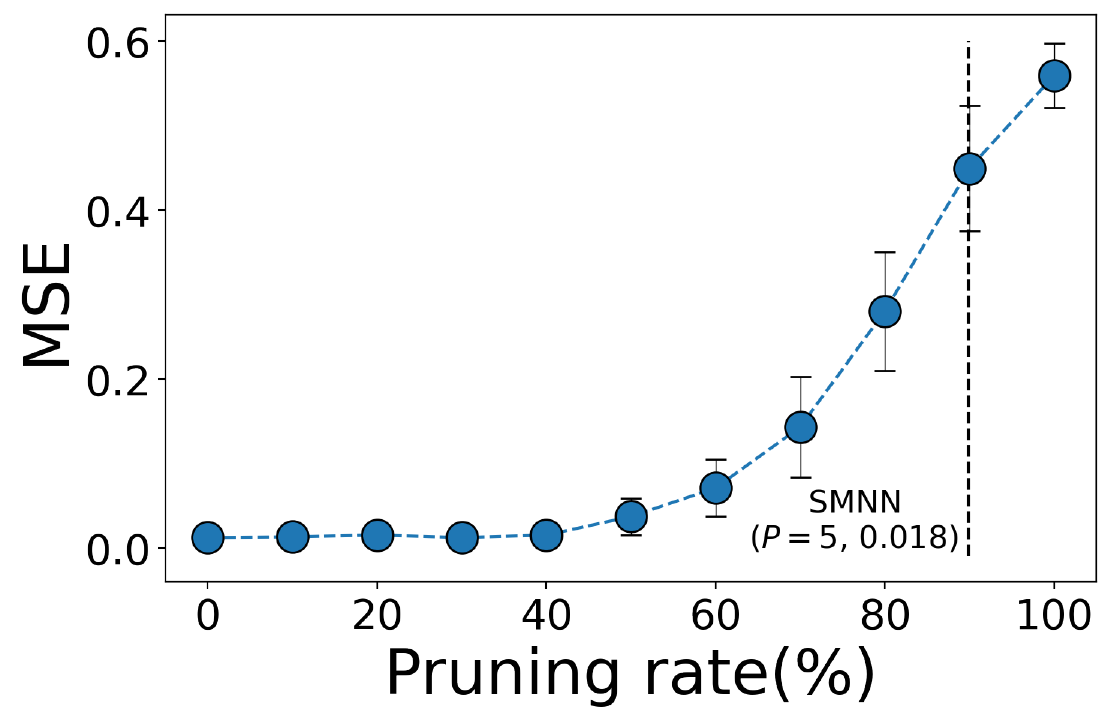}
}
  \caption{Test performance in pruning experiments on trained SNNs ($N=100$) where the weights are set to zero if its absolute value falls below a threshold.  (a) Test accuracy for the MNIST classification task. (b) Test mean square error for the contextual integration task. The vertical line marks the model complexity of SMNN [$(1-(2NP+P)/N^2)$$\times100\%$], and the corresponding test performances are specified in the brackets.
}
  \label{prune2}
\end{figure}

Finally, we also carry out the pruning experiment on trained SNNs where the mode decomposition is not used. As shown in Figure~\ref{prune2}, the SMNN is able to maintain a better accuracy with fewer parameters compared to a well-trained SNN subject to an equivalent pruning after training.

\section{Conclusion}
In this work, we propose a spiking mode-based neural network framework for training various computational tasks. This framework is based on mode decomposition learning inspired from the Hopfield network and multi-layered perceptron training~\cite{Jiang-2021,Li-2023}. From the SMNN learning rule, we can adapt the mode size to the task difficulty, and retrieve the power-law behavior of the importance scores, and furthermore, the high dimensional recurrent activity can be projected to the low-dimensional mode space with a few leading modes, derived from the power-law behavior. Using a few modes, we can speed up the training of recurrent spiking networks, thereby making a large scale of spike-based computation possible in practice. Further extension of our work allows us to treat more biological networks, e.g., considering excitatory-inhibitory networks, sparsity of network connections, and sequence memory from network activity, etc. It would be very interesting to see whether a biological plausible rule can be derived for a sparse sign-constrained  spiking network, which we leave for future works.

To conclude, 
our work provides a fast, interpretable and biologically plausible framework for analyzing neuroscience experiments and designing brain-like
computation circuits, and in particular, the derived plasticity rule in high dimensional spiking dynamics is linked to intriguing physics pictures of
 attractors, power-law behavior, which would help to elucidate inner workings of high dimensional brain dynamics---a central topic of non-equilibrium physics.

\section{Acknowledgments}
We thank Yang Zhao for an earlier involvement of this project, especially the derivation of the discretization of continuous neural and synaptic dynamics,
 and  Yuhao Li for helpful discussions. We also thank Liru Zhang and Ruoran Bai for careful reading of this manuscript.
This research was supported by the National Natural Science Foundation of China for
Grant Number 12122515 (H.H.), and Guangdong Provincial Key Laboratory of Magnetoelectric Physics and Devices (No. 2022B1212010008), 
and Guangdong Basic and Applied Basic Research Foundation (Grant No. 2023B1515040023).  

\appendix
\section{Discretization of continuous neural and synaptic dynamics}\label{app-1}
The interested continuous dynamics can be written in the following form,
\begin{equation}
\tau\frac{\mathrm{d}x}{\mathrm{d}t}=-x(t)+y(t),
\end{equation}
where $y(t)$ is a time-dependent driving term. This first-order linear differential equation has a solution,
\begin{equation}
x(t)=\frac{1}{\tau}e^{-t/\tau}\int_{0}^ty(s)e^{s/\tau}ds,
\end{equation}
which implies that $x(0)=0$. Next, we assume the discretization step size $\Delta t$ is a small quantity. We then have
\begin{equation}\label{discretization}
\begin{split}
x(t+\Delta t)&=\frac{1}{\tau}e^{-\frac{t+\Delta t}{\tau}}\int_0^{t+\Delta t}y(s)e^{s/\tau}ds\\
&=\lambda_\tau x(t)+\frac{1}{\tau}e^{-\frac{t+\Delta t}{\tau}}\int_{t}^{t+\Delta t}y(s)e^{s/\tau}ds\\
&=\lambda_\tau x(t)+\frac{\lambda_\tau}{\tau}\int_{0}^{\Delta t}y(t+s)e^{s/\tau}ds\\
&\simeq\lambda_\tau x(t)+\frac{\lambda_\tau}{\tau}y(t)\int_0^{\Delta t}e^{s/\tau}ds\\
&=\lambda_\tau x(t)+(1-\lambda_\tau)y(t),
\end{split}
\end{equation}
where we change the integral variable in the third equality, and the approximation in the fourth line holds when $\Delta t$ is close to zero, and we define $\lambda_\tau=e^{-\Delta t/\tau}$. Note that in our simulations, $\tau_r$ is relatively small, and therefore we neglect the corresponding decay factor in the second term of the last line in Eq.~\eqref{discretization}. A full set of discrete dynamics is given in Eq.~\eqref{smnndyn}. If a single exponential synaptic filter is considered, the above discretization equation [Eq.~\eqref{discretization}] can be directly used, and $\lambda_\tau=e^{-\Delta t/\tau_{\rm syn}}$, where $\tau_{\rm syn}$ is the synaptic decay time constant as used in a recent work~\cite{Zenke-2022}.

\section{Derivation of mode-based learning rules}\label{app-2}
In this section, we provide details to derive the spiking mode-based learning rules. First, 
the discrete update rules for the dynamics are given below,
\begin{equation}\label{smnndyn}
\begin{aligned}
\ra(t+1) &= \lambda_d\ra(t)+(1-\lambda_d)\mathbf{h}(t+1),\\
\mathbf{h}(t+1) &= \lambda_r\mathbf{h}(t)+\Sa(t+1),\\
\U(t+1) &= \Big( \lambda_{\rm mem}\U(t)+(1-\lambda_{\rm mem})\I(t+1))\Big) \odot\Big( \mathcal{I}-\Sa(t) \Big),\\
\I(t+1) &= \bm{W}^{\rm rec}\ra(t)+\bm{W}^{\rm in}\mathbf{u}(t),
\end{aligned}
\end{equation}
where $t$ is a discrete time step (or in the unit of $\Delta t$), and $\lambda_{\rm mem}=e^{-\Delta t/\tau_{\rm mem}}$. With the loss function $\mathcal{L}$, we define the following error gradients
$
\bm{\mathcal{K}}(t)\equiv\frac{\partial\mathcal{L}}{\partial\ra(t)}
$, and by using the chain rule, we immediately have the following results.
\paragraph{case 1} $t=T$:
\begin{equation}
\begin{aligned}
\frac{\partial\mathcal{L}}{\partial\ra(T)}&=\bm{\mathcal{K}}(T),\\
\frac{\partial\mathcal{L}}{\partial\I(T)}&=\bm{\mathcal{K}}(T)\frac{\partial \ra(T)}{\partial\I(T)}.
\end{aligned}
\end{equation}
\paragraph{case 2} $t<T$:
\begin{equation}
\begin{aligned}
\frac{\partial\mathcal{L}}{\partial\ra(t)}&=\boldsymbol{\mathcal{G}}(t) +\bm{\mathcal{K}}(t+1)\frac{\partial r(t+1)}{\partial\ra(t)}\\
&=\boldsymbol{\mathcal{G}}(t) +\lambda_d\bm{\mathcal{K}}(t+1),\\
\frac{\partial\mathcal{L}}{\partial\I(t)}&=\frac{\partial\mathcal{L}}{\partial \I(t+1)}\frac{\partial \I(t+1)}{\partial\I(t)}+\frac{\partial\mathcal{L}}{\partial \ra(t)}\frac{\partial \ra(t)}{\partial\I(t)}\\
&= \bm{\mathcal{K}}(t)\frac{\partial \ra(t)}{\partial\I(t)},
\end{aligned}
\end{equation}
where $\boldsymbol{\mathcal{G}}(t)$ is the explicit differentiation of $\mathcal{L}$ with respect to $\ra(t)$. Using the surrogate gradient of the step function, we have
\begin{equation}
\begin{aligned}
\Sa(t)&=\Theta(\U(t)-U_{\rm thr}\mathcal{I})\approx\frac{\U(t)-U_{\rm thr}\mathcal{I}}{1+k|\U(t)-U_{\rm thr}\mathcal{I}|},\\
\frac{\partial \Sa(t)}{\partial \U(t)}&={\operatorname{diag}}\Big(\frac{1}{(1+k|\U(t)-U_{\rm thr}\mathcal{I}|)^2}\Big),
\end{aligned} 
\end{equation}
where $\mathcal{I}$ represents an all-one vector, and $|\mathbf{a}|$ for a vector $\mathbf{a}$ equals to $|a_i|$ when the $i$-th component is considered.
As a result,
we can easily get
\begin{equation}\label{drdI}
\begin{aligned}
\frac{\partial \ra(t)}{\partial\I(t)}&=\frac{\partial \ra(t)}{\partial\mathbf{h}(t)}\frac{\partial \mathbf{h}(t)}{\partial\Sa(t)}\frac{\partial \Sa(t)}{\partial\U(t)}\frac{\partial \U(t)}{\partial\I(t)}\\
&=(1-\lambda_d)(1-\lambda_{\rm mem}){\operatorname{diag}}\Big(\frac{1}{(1+k|\U(t)-U_{\rm thr}\mathcal{I}|)^2}\Big).    
\end{aligned}
\end{equation}

Therefore, we can update the mode component and the connectivity importance as follows,
\begin{equation}
\begin{aligned}
\frac{\partial\mathcal{L}}{\partial \bm{\xi}^{\rm in}}&=\sum_{t=1}^T\frac{\partial\mathcal{L}}{\partial \I(t)}\frac{\partial\I(t)}{\partial \bm{W}^{\rm rec}}\frac{\partial\bm{W}^{\rm rec}}{\partial \bm{\xi}^{\rm in}}=\sum_{t=1}^T\frac{\partial\mathcal{L}}{\partial \I(t)}\ra(t-1)\bm{\xi}^{\rm out}\bm{\Sigma},\\
\frac{\partial\mathcal{L}}{\partial \lambda_\mu}&=\sum_{t=1}^T\frac{\partial\mathcal{L}}{\partial \I(t)}\frac{\partial\I(t)}{\partial \bm{W}^{\rm rec}}\frac{\partial\bm{W}^{\rm rec}}{\partial \lambda_{\mu}}=\sum_{t=1}^T\frac{\partial\mathcal{L}}{\partial \I(t)}\ra(t-1)\bm{\xi}^{\rm in}_\mu(\bm{\xi}^{\rm out}_\mu)^\top,\\
\frac{\partial\mathcal{L}}{\partial \bm{\xi}^{\rm out}}&=\sum_{t=1}^T\frac{\partial\mathcal{L}}{\partial \I(t)}\frac{\partial\I(t)}{\partial \bm{W}^{\rm rec}}\frac{\partial\bm{W}^{\rm rec}}{\partial \bm{\xi}^{\rm out}}=\sum_{t=1}^T\frac{\partial\mathcal{L}}{\partial \I(t)}\ra(t-1)\bm{\xi}^{\rm in}\bm{\Sigma},
\end{aligned}
\end{equation}
where $\bm{\xi}_\mu^{\rm{in}/\rm{out}}\in\mathbb{R}^N$ ($\mu$-th column of $\bm{\xi}^{\rm{in}/\rm{out}}$).

For the handwritten digit classification task, $a_i = \max\limits_{t}\{r_i(t)\}=r_i(t_i^m),\mathbf{o} = {\rm{softmax}}(\bm{W}^{\rm out}\mathbf{a})$ and $\mathcal{L}=-\mathbf{z}^T\ln\mathbf{o}$, where $t_i^m$ is the time when the maximal firing rate is reached, and $\mathbf{z}$ is the one-hot target label. We then have the following equations for updating time-dependent error gradients.
\begin{equation}
\begin{aligned}
\mathcal{K}_{i}(T)&=\frac{\partial\mathcal{L}}{\partial a_i}\sum_j\frac{\partial a_i}{\partial r_j(T)}\\
&=\Big((\mathbf{o}-\mathbf{z})\bm{W}^{\rm out}\Big)_i\sum_j\frac{\partial r_i(t_i^m)}{\partial r_j(T)}\\
&=\Big((\mathbf{o}-\mathbf{z})\bm{W}^{\rm out}\Big)_i\sum_j\frac{\partial r_i(t_i^m)}{\partial r_j(t_i^m)}\delta(T-t_i^m)\\
&=\Big((\mathbf{o}-\mathbf{z})\bm{W}^{\rm out}\Big)_i\delta(T-t_i^m),\label{bptta}
\end{aligned}
\end{equation}
and
\begin{equation}
\begin{aligned}
\mathcal{K}_{i}(t)&=\frac{\partial\mathcal{L}}{\partial a_i}\sum_j\frac{\partial a_i}{\partial r_j(t)}+\lambda_d\mathcal{K}_i(t+1)\\
&=\Big((\mathbf{o}-\mathbf{z})\bm{W}^{\rm out}\Big)_i\sum_j\frac{\partial a_i}{\partial r_j(t)}+\lambda_d\mathcal{K}_i(t+1),\forall t<T.\label{bpttb}
\end{aligned}
\end{equation}
The sum in the first term of the last equality in Eq.~\eqref{bpttb} can be explicitly calculated out, i.e.,
\begin{equation}
\begin{aligned}
\sum_j\frac{\partial a_i}{\partial r_j(t)}&=\sum_j\frac{\partial r_i(t_i^m)}{\partial r_j(t)}\\
&=\sum_j\frac{\partial r_i(t_i^m)}{\partial r_j(t_i^m)}\delta(t-t_i^m)+\sum_j\frac{\partial r_i(t_i^m)}{\partial r_j(t_i^m-1)}\delta(t+1-t_i^m)\\
&=\delta(t+1-t_i^m)\sum_j\Big[\delta_{ij}\lambda_d
+\left[\frac{\partial \ra(t_i^m)}{\partial\I(t_i^m)}\right]_{ii}W^{\rm rec}_{ij}\Big]+\delta(t-t_i^m).
\end{aligned}
\end{equation}

For the contextual integration task, $o(t) = \bm{W}^{\rm out}\ra$ and $\mathcal{L} = \sqrt{\sum_{t = 0}^T[z(t)-o(t)]^2}$, 
where $z(t)$ is the target output. In a similar way, one can derive the following error gradients,
\begin{subequations}
\begin{align}
\bm{\mathcal{K}}(T)&=\frac{z(T)-o(T)}{\mathcal{L}}\bm{W}^{\rm out},\\
\bm{\mathcal{K}}(t)&=\frac{z(t)-o(t)}{\mathcal{L}}\bm{W}^{\rm out}+\lambda_d\bm{\mathcal{K}}(t+1),\forall t<T.        
\end{align}
\end{subequations}
The other update equations are the same as above.

\section{Initialization and model parameters}\label{app-3}
\label{init}
The initialization scheme is given in Table~\ref{tab1}, where the constructed recurrent weights should be multiplied by a factor of $\frac{1}{\sqrt{PN}}$~\cite{Li-2023}.
The hyper parameters used in the neural dynamics equations are given in Table~\ref{tab2}.

\begin{table}[bt]
\centering
\caption{\label{tab1}Parameter initialization}
\begin{tabular}{lll}
\hline
  Parameter & Initial distribution & Description\\
  \hline
  $\bm{W}^{\rm in}$ & $\mathcal{N}(0,1/N_{\rm in})$ & Input weight\\
      $\bm{\xi}^{\rm in}$ & $\mathcal{N}(0,1)$ & Input mode\\
      $\bm{\xi}^{\rm out}$ & $\mathcal{N}(0,1)$ & Output mode\\
      $\lambda_\mu$ & $\mathcal{N}(0,1)$ & Connectivity importance\\
      $\bm{W}^{\rm out}$ & $\mathcal{N}(0,1/N)$ & Readout weight\\
\hline
\end{tabular}
\end{table}

\begin{table}[bt]
\centering
\caption{\label{tab2}Model parameters}
\begin{tabular}{lll}
\hline
  Parameter & Value & Description\\
\hline
   $\Delta t$ & 0.2\,ms & Discretization step size\\
      $\tau_{\rm mem}$ & 20\,ms & Membrane time constant\\
      $\tau_{r}$ & 2\,ms & Synaptic rise time\\
      $\tau_{d}$ & 30\,ms & Synaptic decay time\\
      $U_{\rm thr}$ & 1 & Spiking threshold\\
       $U_{\rm res}$ & 0 & Resting potential\\
      $t_{\rm ref}$ & 2\,ms & Refractory period\\
      $k$ & 25 & Steepness parameter\\
\hline
\end{tabular}
\end{table}


\begin{thebibliography}{10}

\bibitem{ND-2014}
Wulfram Gerstner, Werner~M. Kistler, Richard Naud, and Liam Paninski.
\newblock {\em Neuronal Dynamics: From Single Neurons to Networks and Models of
  Cognition}.
\newblock Cambridge University Press, United Kingdom, 2014.

\bibitem{Nat-2019}
Kaushik Roy, Akhilesh Jaiswal, and Priyadarshini Panda.
\newblock Towards spike-based machine intelligence with neuromorphic computing.
\newblock {\em Nature}, 575(7784):607--617, 2019.

\bibitem{Werbos-1990}
Paul~J. Werbos.
\newblock Backpropagation through time: What it does and how to do it.
\newblock {\em Proc. IEEE}, 78:1550--1560, 1990.

\bibitem{Elman-1990}
Jeffrey~L. Elman.
\newblock Finding structure in time.
\newblock {\em Cognitive Science}, 14(2):179--211, 1990.

\bibitem{Huang-2022}
Haiping Huang.
\newblock {\em Statistical Mechanics of Neural Networks}.
\newblock Springer, Singapore, 2022.

\bibitem{Zenke-2019}
Emre~O. Neftci, Hesham Mostafa, and Friedemann Zenke.
\newblock Surrogate gradient learning in spiking neural networks: Bringing the
  power of gradient-based optimization to spiking neural networks.
\newblock {\em IEEE Signal Processing Magazine}, 36(6):51--63, 2019.

\bibitem{Abbott-2009}
David Sussillo and Larry~F Abbott.
\newblock Generating coherent patterns of activity from chaotic neural
  networks.
\newblock {\em Neuron}, 63(4):544--557, 2009.

\bibitem{Kappen-2016}
Dominik Thalmeier, Marvin Uhlmann, Hilbert~J Kappen, and Raoul-Martin
  Memmesheimer.
\newblock Learning universal computations with spikes.
\newblock {\em PLoS computational biology}, 12(6):e1004895, 2016.

\bibitem{Force-2017}
Wilten Nicola and Claudia Clopath.
\newblock Supervised learning in spiking neural networks with force training.
\newblock {\em Nature communications}, 8(1):2208, 2017.

\bibitem{Kim-2018}
Christopher~M Kim and Carson~C Chow.
\newblock Learning recurrent dynamics in spiking networks.
\newblock {\em Elife}, 7:e37124, 2018.

\bibitem{Kim-2019}
Robert Kim, Yinghao Li, and Terrence~J Sejnowski.
\newblock Simple framework for constructing functional spiking recurrent neural
  networks.
\newblock {\em Proceedings of the national academy of sciences},
  116(45):22811--22820, 2019.

\bibitem{Abbott-2016b}
Brian DePasquale, Mark~M Churchland, and LF~Abbott.
\newblock Using firing-rate dynamics to train recurrent networks of spiking
  model neurons.
\newblock {\em arXiv:1601.07620}, 2016.

\bibitem{Abbott-2016a}
Larry~F Abbott, Brian DePasquale, and Raoul-Martin Memmesheimer.
\newblock Building functional networks of spiking model neurons.
\newblock {\em Nature neuroscience}, 19(3):350--355, 2016.

\bibitem{Zenke-2018}
Friedemann Zenke and Surya Ganguli.
\newblock Superspike: Supervised learning in multilayer spiking neural
  networks.
\newblock {\em Neural Computation}, 30:1--28, 2018.

\bibitem{Ostojic-2021}
Mehrdad Jazayeri and Srdjan Ostojic.
\newblock Interpreting neural computations by examining intrinsic and embedding
  dimensionality of neural activity.
\newblock {\em Current Opinion in Neurobiology}, 70:113--120, 2021.

\bibitem{Alexis-2022}
Alexis Dubreuil, Adrian Valente, Manuel Beiran, Francesca Mastrogiuseppe, and
  Srdjan Ostojic.
\newblock The role of population structure in computations through neural
  dynamics.
\newblock {\em Nature Neuroscience}, 25(6):783--794, 2022.

\bibitem{Ostojic-2023}
Manuel Beiran, Nicolas Meirhaeghe, Hansem Sohn, Mehrdad Jazayeri, and Srdjan
  Ostojic.
\newblock Parametric control of flexible timing through low-dimensional neural
  manifolds.
\newblock {\em Neuron}, 111(5):739--753, 2023.

\bibitem{Brian-2023}
Brian DePasquale, David Sussillo, LF~Abbott, and Mark~M Churchland.
\newblock The centrality of population-level factors to network computation is
  demonstrated by a versatile approach for training spiking networks.
\newblock {\em Neuron}, 111(5):631--649, 2023.

\bibitem{Jaz-2020}
Eli Pollock and Mehrdad Jazayeri.
\newblock Engineering recurrent neural networks from task-relevant manifolds
  and dynamics.
\newblock {\em PLOS Computational Biology}, 16(8):1--23, 2020.

\bibitem{Sussillo-2015}
David Sussillo, Mark~M Churchland, Matthew~T Kaufman, and Krishna~V Shenoy.
\newblock A neural network that finds a naturalistic solution for the
  production of muscle activity.
\newblock {\em Nature Neuroscience}, 18(7):1025--1033, 2015.

\bibitem{Ostojic-2018}
Francesca Mastrogiuseppe and Srdjan Ostojic.
\newblock Linking connectivity, dynamics, and computations in low-rank
  recurrent neural networks.
\newblock {\em Neuron}, 99(3):609--623, 2018.

\bibitem{Ostojic-2023b}
Ljubica Cimesa, Lazar Ciric, and Srdjan Ostojic.
\newblock Geometry of population activity in spiking networks with low-rank
  structure.
\newblock {\em PLOS Computational Biology}, 19(8):1--34, 2023.

\bibitem{CK-2023}
William~F. Podlaski and Christian~K. Machens.
\newblock Approximating nonlinear functions with latent boundaries in low-rank
  excitatory-inhibitory spiking networks.
\newblock {\em arXiv:2307.09334}, 2023.

\bibitem{Jiang-2021}
Zijian Jiang, Jianwen Zhou, Tianqi Hou, K.~Y.~Michael Wong, and Haiping Huang.
\newblock Associative memory model with arbitrary hebbian length.
\newblock {\em Phys. Rev. E}, 104:064306, 2021.

\bibitem{Jiang-2023}
Zijian Jiang, Ziming Chen, Tianqi Hou, and Haiping Huang.
\newblock Spectrum of non-hermitian deep-hebbian neural networks.
\newblock {\em Phys. Rev. Res.}, 5:013090, 2023.

\bibitem{Li-2023}
Chan Li and Haiping Huang.
\newblock Emergence of hierarchical modes from deep learning.
\newblock {\em Phys. Rev. Res.}, 5:L022011, 2023.

\bibitem{Zenke-2022}
Julian Rossbroich, Julia Gygax, and Friedemann Zenke.
\newblock Fluctuation-driven initialization for spiking neural network
  training.
\newblock {\em Neuromorphic Computing and Engineering}, 2(4):044016, 2022.

\bibitem{Mante-2013}
Valerio Mante, David Sussillo, Krishna~V Shenoy, and William~T Newsome.
\newblock Context-dependent computation by recurrent dynamics in prefrontal
  cortex.
\newblock {\em Nature}, 503(7474):78--84, 2013.

\bibitem{XJ-2016}
H~Francis Song, Guangyu~R Yang, and Xiao-Jing Wang.
\newblock Training excitatory-inhibitory recurrent neural networks for
  cognitive tasks: a simple and flexible framework.
\newblock {\em PLoS computational biology}, 12(2):e1004792, 2016.

\bibitem{Hopfield-1982}
J~J Hopfield.
\newblock Neural networks and physical systems with emergent collective
  computational abilities.
\newblock {\em Proceedings of the National Academy of Sciences},
  79(8):2554--2558, 1982.

\bibitem{Cramer-2019}
Benjamin Cramer, Yannik Stradmann, Johannes Schemmel, and Friedemann Zenke.
\newblock The heidelberg spiking data sets for the systematic evaluation of
  spiking neural networks.
\newblock {\em IEEE Transactions on Neural Networks and Learning Systems},
  33(7):2744--2757, 2022.

\bibitem{mnist}
Yann LeCun.
\newblock The MNIST database of handwritten digits, retrieved from
  http://yann.lecun.com/exdb/mnist., 1998.

\bibitem{Miconi-2017}
Thomas Miconi.
\newblock Biologically plausible learning in recurrent neural networks
  reproduces neural dynamics observed during cognitive tasks.
\newblock {\em Elife}, 6:e20899, 2017.

\bibitem{adam}
Diederik~P. Kingma and Jimmy Ba.
\newblock Adam: A method for stochastic optimization.
\newblock {\em arXiv:1412.6980}, 2014.

\bibitem{ZH-2023}
Zhanghan Lin and Haiping Huang.
\newblock https://github.com/LinZhanghan/SMNN, 2023.

\bibitem{Pos-2009}
Bruno~B. Averbeck.
\newblock Poisson or not poisson: Differences in spike train statistics between
  parietal cortical areas.
\newblock {\em Neuron}, 62(3):310--311, 2009.

\end{thebibliography}

\end{document}